\documentclass[12pt]{article}

\begin{document}

\setcounter{page}{0} \thispagestyle{empty}


\vskip 1cm

\begin{center}
{\Large {\bf 
On the highest transcendentality  
in ${\mathcal N}=4$ SUSY
}
\footnote{This work was supported by the Marie Curie and
RFBR grants 04-02-17094,
RSGSS-5788.2006.2}} \\[0pt]
\vspace{1.5cm} {\large \ A.~V.~Kotikov$^a$ and L.~N.~Lipatov$^{b,c}$
} \\[0pt]
\vspace{1cm} {$^a$ \em Bogoliubov Laboratory of Theoretical Physics \\[0pt]
Joint Institute for Nuclear Research\\[0pt]
141980 Dubna, Russia }\\[0pt]
\vspace{0.5cm} {$^b$ \em Theoretical Physics Department\\[0pt]
Petersburg Nuclear Physics Institute\\[0pt]
Orlova Roscha, Gatchina\\[0pt]
188300, St. Petersburg, Russia }\\[0pt]
\vspace{0.5cm} {$^c$ \em II. Institut f\" ur Theoretische Physik, Universit\" at Hamburg\\[0pt]
Luruper Chaussee 149, 22761 Hamburg, Germany }\\[0pt]
\end{center}

\vspace{1.5cm}\noindent

\abstract{
\noindent We investigate the
Eden-Staudacher equation for the anomalous dimension of the
twist-2 operators at the large spin $s$ in the $N=4$ super-symmetric gauge 
theory.
This equation is reduced to a set of linear algebraic equations with the kernel 
calculated analytically. We prove that in perturbation theory the 
anomalous dimension is a sum of products of  
the Euler functions $\zeta(k)$ having the property of
the maximal transcendentality with the coefficients being integer numbers.
The radius of convergency of the perturbation theory is found.
It is shown, that at $g=\infty$ the kernel has an essential singularity.
The analytic properties of the solution of the Eden-Staudacher equation
are investigated. In particular for the case of the strong coupling constant
the solution has an essential singularity on the second sheet of the 
variable $j$ appearing in its Laplace transformation. Similar results 
are derived also for the Beisert-Eden-Staudacher equation which includes 
the contribution from the phase related to the
crossing symmetry of the underlying $S$-matrix. We show, that
its singular solution at large coupling constants reproduces 
the anomalous dimension predicted from the string side of the 
AdS/CFT correspondence.}

\section{Introduction}

The anomalous dimension matrices $\gamma $ and $\tilde{\gamma}$ of the twist-2 Wilson 
operators in Quantum Chromodynamics (QCD) govern the
Bjorken scaling violation for structure functions respectively for the non-polarized
and polarized cases. These quantities calculated in terms of the
perturbative expansion in $a_s=\alpha_s/(4\pi)$ 
are related with the Mellin transformation 
\begin{eqnarray}
\gamma _{ab}(j) &=& \int_{0}^{1}dx\,\,x^{j-1} W_{b\rightarrow a}(x)
~=~ \gamma^{(0)} _{ab}(j) a_s + \gamma^{(1)} _{ab}(j) a_s^2+
\gamma^{(2)} _{ab}(j) a_s^3 + O(a_s^4),~~
\nonumber \\
\tilde{\gamma} _{ab}(j) &=& \int_{0}^{1}dx\,\,x^{j-1}
\tilde{W}_{b\rightarrow a}(x)
~=~ \tilde{\gamma}^{(0)} _{ab}(j) a_s + \tilde{\gamma}^{(1)} _{ab}(j) a_s^2+
\tilde{\gamma}^{(2)} _{ab}(j) a_s^3 + O(a_s^4)
\end{eqnarray}
to the splitting kernels $W_{b\rightarrow a}(x)$ and
$\tilde{W}_{b\rightarrow a}(x)$ for the
Dokshitzer-Gribov-Lipatov-Altarelli-Parisi (DGLAP) equation~\cite{DGLAP}
describing the evolution of the parton densities $f_{a}(x,Q^{2})$ and
$\tilde{f}_{a}(x,Q^{2})$
(hereafter $a=\lambda,\,g,\,\phi$ for
the spinor, vector and scalar particles)
\begin{eqnarray}
\frac{d}{d\ln {Q^{2}}}f_{a}(x,Q^{2}) &=& \int_{x}^{1}\frac{dy}{y}
\sum_{b}W_{b\rightarrow a}(x/y)\,f_{b}(y,Q^{2}) \, ,
\nonumber \\
\frac{d}{d\ln {Q^{2}}}\tilde{f}_{a}(x,Q^{2}) &=& \int_{x}^{1}\frac{dy}{y}
\sum_{b}\tilde{W}_{b\rightarrow a}(x/y)\,\tilde{f}_{b}(y,Q^{2}) \,.
\label{DGLAP}
\end{eqnarray}
The scalar particles appear in the supersymmetric models but in
the spin-dependent case their contribution is absent $a=\lambda,\,g$.
The anomalous dimensions and splitting kernels in QCD are known up to the
next-to-next-to-leading order (NNLO) of the perturbation theory
~\cite{LONLOAD,VMV}.

The QCD expressions for anomalous dimensions can be transformed to the case
of the
${\mathcal N}$-extended Supersymmetric Yang-Mills Models (SYM)
if one will use for the Casimir operators $C_{A},C_{F},T_{f}$ the following
values $C_{A}=C_{F}=N_{c}$, $T_{f}n_f={\mathcal N}N_{c}/2$.
For ${\mathcal N}\!\!=\!\!2$ and ${\mathcal N}\!\!=\!\!4$ 
the anomalous dimensions of the Wilson operators get also
additional contributions coming from diagrams with scalar particles~\cite{KL}.
These anomalous dimensions were calculated in the next-to-leading order (NLO)~\cite{KoLiVe}
for ${\mathcal N}=4$ SYM.

It turns out, that the expressions
for eigenvalues of the anomalous dimension matrix in ${\mathcal N}=4$ 
SYM can be derived directly
from the QCD anomalous dimensions without tedious calculations by using a
number of plausible arguments. The method elaborated in Ref.~\cite{KL} for
this purpose (called {\it the maximal transcendentality principle})
is based on special properties of
the integral kernel for
the Balitsky-Fadin-Kuraev-Lipatov (BFKL) equation~\cite{BFKL}-\cite{KL00}
and on an interesting relation between the BFKL and DGLAP equations 
in this model (see~\cite{KL}).
Really it was assumed, that the coefficients of the perturbation theory for the
BFKL kernel and for the universal anomalous dimension should be
linear combinations of the most complicated special functions which could 
appear in each given order. 
The results~\cite{KL} for the
anomalous dimension obtained with the use of  the maximal transcedentality 
hypothesis were checked 
by direct calculations in Ref.~\cite{KoLiVe}.
Using the three-loop expressions for  anomalous dimensions in QCD~\cite{VMV}
and this hypothesis the eigenvalues of the anomalous 
dimension matrix for the ${\mathcal N}=4$ SYM in the NNLO
approximation were derived~\cite{KLOV}. The direct verification of the obtained result for 
the case of the Konishi operator was done in Ref. \cite{Konish}. Also the independent  
calculation of the anomalous dimension for the twist-2 operator at the large spin $s$
gives the same correction in three loops~\cite{Bern1}. Recently the four-loop contribution at 
$s\rightarrow \infty$ was computed and the KLV procedure~\cite{KoLiVe} was used 
for the resummation of the perturbation theory~\cite{BCDKS}.  The perturbative results are 
important for the verification of the
AdS/CFT correspondence ~\cite{AdS-CFT} and for checking the integrability of this model. 
Historically the integrability 
of the Yang-Mills theory at large energies appeared
in the context of the solution of the Bartels-Kwiecinski-Prascalowich equation \cite{BKP} in the 
multi-colour 
limit~\cite{integrBF}. The effective hamiltonian of this equation coincides with the
local hamiltonian of the Heisenberg spin model~\cite{Heis}. Later it was shown, 
that the equations for the anomalous dimensions of the
so-called quasi-partonic operators 
~\cite{QPO} in the leading logarithmic approximation
are also integrable and related to another Heisenberg spin model, but only in 
${\mathcal N}=4$
multi-colour SUSY~\cite{integrDG}.  Partly these remarkable properties remain also in
QCD for a restricted class of 
quasi-partonic operators~\cite{BDMKB}. After the discovery of
the AdS/CFT correspondence the integrability of the 
${\mathcal N}=4$ SYM
was established
in many loops in the weak and strong coupling regimes (see, for example, Ref.~\cite{BeiSta}
and references therein).
Several months ago in an interesting paper Eden and Staudacher (ES) derived an integral 
equation for the function which governs the behavior of the anomalous dimension for
the twist-2 operators at large spins $s$ in all orders of the perturbation theory~\cite{EdenSt} (see also \cite{Bel}). 
Its modification, which takes into account in the $S$-matrix  an additional phase factor restoring 
the correspondence between the conformal field theory and the superstring model, was
suggested recently by Beisert, Eden and Staudacher (BES)~\cite{BES}. Our paper is devoted
to an analytic solution of the ES and BES equations. Partly our results were presented by one 
of authors (L.L.) on the Potsdam Workshop~\cite{LipPots}. The numerical solutions of the BES 
equation for all coupling constants were discussed recently 
in the papers of Bern with 
collaborators~\cite{BCDKS} and 
by Klebanov et al.~\cite{Klebanov} in the framework of the equivalent set of 
linear algebraic equations discussed by one of us in the Potsdam talk~\cite{LipPots}.

Our paper is organized as follows. In Section 2 the ES equation is simplified with the use of 
the inverse Laplace representation. Section 3 contains the derivation of a set of linear
algebraic equations for the coefficients $a_{n,\epsilon}$ appearing in the expansion of its
solution in terms of some special functions. The kernel for these equations is calculated 
in an explicit form and its analytic properties
in the coupling constant plane are investigated. In Section 4 we study
the anomalous dimension in the perturbation theory, verify its maximal transcedentality 
property and prove the Eden-Staudacher hypothesis~\cite{EdenSt} about the integer coefficients 
in the sum
of products of the corresponding $\zeta$-functions.
Section 5 is devoted to the investigation of analytical properties of the solution 
$\phi (j)$ of the ES equation in the Laplace variable $j$ for arbitrary complex constants.
The obtained results give us a possibility to find the functional relations 
for $\phi (j)$ equivalent to the ES equation and to calculate the asymptotic behavior of the coefficients 
$a_{n, \epsilon}$. In  Section 6 we study analytic properties of the solution 
of the ES equation at large coupling constants. In particular its behavior near an 
essential singularity on the second sheet of the $j$-plane is investigated. Using some plausible arguments
we calculate from this result the anomalous dimension at large coupling constants. It oscillates around 
the value predicted from the string theory. Sections 7 and 8 contain
a similar analysis for the Beisert-Eden-Staudacher equation. It turns out, that in this case the
behavior of the singular anomalous dimension at large coupling constants is stabilized and coincides
with the result predicted from the string side of the AdS/CFT correspondence. In Conclusion we discuss
obtained results and unsolved problems.

In Appendix A the duality relation for the dressing phase factor, proposed in \cite{BES} and entering
in the BES equation, is proved. Appendix B contains an independent derivation of the 
linear sets of equations, obtained in Sections 3 and 7. 
In Appendices C and D their simplified derivation is done for an important case of large couplings.

\section{ES equation in the inverse Laplace
representation}

One can write the anomalous dimension $\Delta -s$ of the twist-2 operators in
the $N=4$ super-symmetric gauge theory at the large Lorentz spin $s \rightarrow \infty$ in the form 
\begin{equation}
\Delta -s =\gamma \,\ln s \,,
\end{equation}
where the coefficient $\gamma$ in the t'Hooft limit depends only of the coupling constant $g$.
Eden and Staudacher obtained for it the expression~\cite{EdenSt}
\begin{equation}
\gamma = 8 \,g^2\, \sigma (0)=4\,g\, \sqrt{2}\,f(0)
\label{def}
\end{equation}
in terms of the function
\begin{equation}
\sigma (t) = \epsilon \,f(x)\,\,,\,\,\,t=\epsilon \,x\,\,,\,\,\,\epsilon =%
\frac{1}{g\,\sqrt{2}}\, .
\label{def1}
\end{equation}
This function satisfies the ES
equation~\cite{EdenSt}
\begin{equation}
\epsilon \,f(x)=\frac{t}{e^t-1}\left(\frac{J_1(x)}{x}-\int _0^\infty
dx^{\prime}\,K(x,x^{\prime})\,f(x^{\prime})\right)
\label{Eq1}
\end{equation}
with the integral kernel
\begin{equation}
K(x,y)=\frac{J_1(x)\,J_0(y)-J_1(y)\,J_0(x)}{x-y}\,,
\label{Eq2}
\end{equation}
where $J_n(x)$ are the Bessel functions
\begin{equation}
J_n(x)=\sum _{k=0}^{\infty}\frac{(-1)^k}{k!\,(k+n)!}\,\left(\frac{x}{2}%
\right)^{2k+n},\,\, J_1(x)=-J_0^{\prime}(x)\,.
\label{Eq3}
\end{equation}

Using the recurrent relation
\begin{equation}
J_{n-1}(x)=-J_{n+1}(x)+2\,n\,x^{-1}\,J_{n}(x)\,,
\label{Eq4}
\end{equation}
we can write the kernel in a simpler way
\begin{equation}
K(x,y)= \frac{2}{x\,y}\,\sum _{n=1}^{\infty}\,n\,J_n(x)\,J_n(y) \,.
\label{Eq5}
\end{equation}

Let us present the solution in the form of the Laplace
integral
\begin{equation}
f(x)= \int _{-i\infty}^{i\infty}\frac{d\,j}{2\pi \,i}\,e^{x\,j}\,\phi (j)\,,
\label{Mellin}
\end{equation}
where $\phi (j)$ is the function analytic in the right semi-plane of the $j$%
-plane and decreasing at $j\rightarrow \infty$ as follows
\begin{equation}
\lim _{j\rightarrow \infty}\phi (j)=\frac{c(\epsilon)}{j}\,.
\label{jlim}
\end{equation}
The residue $c(\epsilon)$ is related directly to the anomalous dimension
\begin{equation}
\gamma = \frac{4}{\epsilon} \,c(\epsilon )\,,
\label{anomjlim}
\end{equation}

Using the Laplace
transformation
\begin{equation}
\phi (j)=\int _{0}^{\infty}\,dx\,e^{-x\,j}\,f(x)
\end{equation}
and the following relations
\begin{equation}
\int _{0}^{\infty}\,dx\,e^{-x\,j}\,J_n(x)=
\frac{\left((j^2+1)^{1/2}-j\right)^n}{
(j^2+1)^{1/2}}=
\frac{\left((j^2+1)^{1/2}+j\right)^{-n}}{(j^2+1)^{1/2}},
\end{equation}
\begin{equation}
\int _{0}^{\infty}\,\frac{dx}{x}\,e^{-x\,j}\,J_n(x)= \frac{1}{n}%
\,\left((j^2+1)^{1/2}-j\right)^n= \frac{1}{n}\,\left((j^2+1)^{1/2}+j%
\right)^{-n}\,,
\end{equation}
\begin{equation}
\int _{0}^{\infty}\,dx\,e^{-x\,j}\,(e^{\epsilon \,x}-1)\,f(x)= \phi
(j-\epsilon )-\phi (j)\,,
\end{equation}
we can write the ES equation for $\phi (j)$ in the form
\begin{equation}
\frac{\phi (j-\epsilon )-\phi (j)}{(j^2+1)^{-1/2}}=\frac{1}{%
(j^2+1)^{1/2}+j} -2\, \int _{-i\infty}^{i\infty}\,\frac{dj^{\prime}}{2\,\pi
\,i}\sum _{n=1}^{\infty}\,\phi (j^{\prime})\,\left(\frac{-(j^{\prime
\,2}+1)^{1/2}+j^{\prime}}{(j^2+1)^{1/2}+j}\right)^n\,.
\end{equation}
Here the integration contour in $j^{\prime}$ is assumed to be to the left
from the cut of the function $-\sqrt{j^{\prime \,2}+1}+j'$ at $%
-i<j^{\prime}<i$ and to the right from the singularities of the function $%
\phi (j')$.
The branch of $%
\sqrt{j^{\prime \,2}+1}$ is chosen in such way to make the expression $-%
\sqrt{j^{\prime \,2}+1}+j'$ decreasing at large $j^{\prime }$.

\section{Set of linear algebraic equations}

We search the solution of the ES equation in the form \cite{LipPots}
 \begin{equation}
\phi (j)=\sum _{n=1}^{\infty}\,\phi _{n,\epsilon
}(j)\,\left(\delta _{n,1}- a_{n,\epsilon }\right)\,,
\label{Phi}
\end{equation}
where $\delta _{n,1}$ is the Kroneker symbol and 
\begin{equation}
\phi _{n,\epsilon }(j)=\sum
_{s=1}^{\infty}\,\frac{\left(\sqrt{(j+s\,\epsilon
)^2+1}+j+s\,\epsilon \right)^{-n}}{\sqrt{(j+s\,\epsilon )^2+1}}
\,.
\end{equation}

The anomalous dimension can be expressed in terms of
$a_{1,\epsilon }$ as follows
\begin{equation}
\gamma =\frac{2}{\epsilon ^2}\,\left(1-a_{1,\epsilon }\right)
\label{gamma}
\end{equation}
The coefficients in the linear combination of the functions $\phi
_{n,\epsilon }(j)$ satisfy the set of equations \cite{LipPots}
\begin{equation}
a_{n,\epsilon }=\sum _{n'=1}^{\infty} \,K_{n,n'}(\epsilon )\,
\left(\delta _{n',1}- a_{n',\epsilon }\right)\,,
\label{an}
\end{equation}
where
\begin{equation}
K_{n,n'}(\epsilon )=2\, \int
_{-i\infty}^{i\infty}\,\frac{dj^{\prime}}{2\,\pi \,i}\,\phi
_{n',\epsilon } (j^{\prime})\,\left(-(j^{\prime
\,2}+1)^{1/2}+j^{\prime}\right)^n\,.
\label{Knn}
\end{equation}
The matrix elements $K_{n,n'}(\epsilon )$ can be expressed in
terms of the generalized hypergeometric function \cite{LipPots}. To show it, 
we begin with
the following representation for $a_{n,\epsilon}$
\begin{equation}
a_{n,\epsilon}=2\, \int _{\sigma -i\infty}^{\sigma
+i\infty}\,\frac{dj^{\prime}}{2\,\pi i}\,\phi
(j^{\prime})\left(-(j^{\prime
\,2}+1)^{1/2}+j^{\prime}\right)^n=-2\sum_{k=n}^{\infty}\,b_n(k)\,
\frac{\phi ^{(k-1)}(0)}{(k-1)!}\,,
\end{equation}
where $0> \sigma >-\epsilon$ and $\phi ^{(n-1)}(0)$ is the value 
of the $(n-1)$-derivative of
the function $\phi (j)$ at $j=0$. The coefficients $b_n(k)$ are
defined in terms of the large-$j$ expansion
\begin{equation}
\left(-(j^{2}+1)^{1/2}+j \right)^n=
\sum_{k=n}^{\infty}b_n(k)\,j^{-k}\,.
\end{equation}

These coefficients can be written as follows
\begin{equation}
b_n(k)=\int _L\frac{d\,j}{2\pi i}\,\left(-(j^{2}+1)^{1/2}+j
\right)^n\,j^{k-1}=(-1)^n\int _L\frac{d\,l}{2\pi i}\,2^k\,
(1+l^2)\,\frac{(1-l^2)^{k-1}}{l^{k-n+1}}\,,
\end{equation}
where the integration contour $L$ goes along a circle around the
singularities of the integrand in a anti-clock-wise direction and
we introduced the new integration variable
\[
l=(j^{2}+1)^{1/2}-j\,.
\]
The integral over $l$ can be calculated by
residues
\begin{equation}
b_n(k)=(-1)^{(k+n)/2}\,2^{-k}\, \frac{n\,\Gamma(k)}{\Gamma
(\frac{k-n}{2}+1)\,\Gamma (\frac{k+n}{2}+1)}\,,
\end{equation}
where
\begin{equation}
\frac{k-n}{2}=r\,
\end{equation}
is an integer number (for half-integer values of $r$ one obtains
$b_n(k)=0$).

We can write $a_{n, \epsilon}$ as follows
\begin{equation}
a_{n, \epsilon}=\sum _{r=0}^{\infty}B_n(r)\,\phi ^{(2r+n-1)}(0)\,,
\end{equation}
where
\begin{equation}
B_n(r)=-2\,\frac{b_n(2r+n)}{(2r+n-1)!}=(-1)^{r+n+1}\,2^{-2r-n+1}\,
\frac{n}{ r!\,(r+n)!}\,.
\end{equation}
Therefore one obtains
\begin{equation}
\left(-(j^{2}+1)^{1/2}+j \right)^n=
-\sum_{r=0}^{\infty}\frac{(2r+n-1)!}{2}\,B_n(r)\,j^{-2r-n}\,.
\end{equation}

To calculate the kernel $K_{n,n'}$ appearing in the set of
equations for $a_{n, \epsilon}$ let us use the expansion
\begin{equation}
\frac{\left((j^{2}+1)^{1/2}-j \right)^{n}}{(j^{2}+1)^{1/2}}=
-\frac{1}{n}\,\frac{\partial}{\partial j}\left((j^{2}+1)^{1/2}-j
\right)^n=-(-1)^n\sum_{r=0}^{\infty}\frac{(2r+n)!}{2\,n\,j^{2r+n+1}}
\,B_n(r)\,j^{-2r-n-1}.
\end{equation}
It gives a possibility to express $\phi _{n,\epsilon }(j)$ in
terms of simpler functions
\begin{equation}
\phi _{n,\epsilon
}(j)=(-1)^{n+1}\sum_{r=0}^{\infty}\frac{(2r+n)!}{2\,n}
\,B_n(r)\,\chi _{2r+n+1,\epsilon}(j)\,,\,\,\chi
_{k,\epsilon}(j)=\sum _{s=1}^{\infty}(j+\epsilon \,s)^{-k}\,.
\end{equation}
In particular,
\begin{equation}
\phi ^{(2r+n-1)}_{n',\epsilon }(0)=(-1)^{n+n'}\sum_{r'=0}^{\infty}
\,B_{n'}(r')\,\frac{(2r+n+2r'+n'-1)!}{2\,n'\,\epsilon
^{2r+n+2r'+n'}}\,\zeta (2r+n+2r'+n')\,,
\end{equation}
where the Euler $\zeta$-function is defined as follows
\begin{equation}
\zeta (k)=\sum _{s=1}^{\infty}s^{-k}\,.
\end{equation}

Using above relations we can present the kernel $K_{n,n'}$ as
follows
\[
K_{n,n'}(\epsilon )
\]
\begin{equation}
=(-1)^{n+n'}\sum
_{r=0}^{\infty}B_{n}(r)\sum_{r'=0}^{\infty}
\,B_{n'}(r')\,\frac{(2r+n+2r'+n'-1)!}{2\,n'\,\epsilon
^{2r+n+2r'+n'}}\,\zeta (2r+n+2r'+n')\,.
\end{equation}
It can be written in the form
\begin{equation}
K_{n,n'}(\epsilon )=2n\sum
_{R=0}^{\infty}\frac{(2R+n+n'-1)!}{\epsilon ^{2R+n+n'}}\,\zeta
(2R+n+n')\,C_{n,n',R}\,,
\end{equation}
where $C_{n,n',R}$ is given below
\begin{equation}
C_{n,n',R}=\frac{(-1)^{n+n'}}{4nn'}\sum
_{r=0}^{\infty}B_{n}(r) \,B_{n'}(R-r)\,
=(-1)^R\,
\frac{F(-R,-R-n';n+1;1)}{2^{2R+n+n'}\,n!\,R!\,(R+n')!}\,.
\end{equation}
Here $F(a,b;c;x)$ is the hypergeometric function which can be calculated
at $x=1$. Therefore one obtains
\begin{equation}
C_{n,n',R}=(-1)^R\,\frac{2^{-2R-n-n'}\,(2R+n+n')!}{R!\, (R+n)!\,
(R+n')!\,(R+n+n')!}\,.
\end{equation}
As a result, we have the following expression for the kernel \cite{LipPots}
\begin{equation}
K_{n,n'}(\epsilon )=2n\sum
_{R=0}^{\infty}(-1)^R\,\frac{2^{-2R-n-n'}}{\epsilon
^{2R+n+n'}}\,\zeta (2R+n+n')\,\frac{
(2R+n+n'-1)!\,(2R+n+n')!}{R!\, (R+n)!\, (R+n')!\,(R+n+n')!}\,.
\label{Eq37}
\end{equation}

The perturbation series for $K_{n,n'}(\epsilon )$
has a finite radius of convergence due to the singularity at
$\epsilon ^{-2}=-1/4$. But it can be analytically
continued around this singularity with the use of the contour integral
representation and known relations for the $\Gamma$-functions
\[
K_{n,n'}(\epsilon )=
\]
\begin{equation}
\frac{n}{\pi}\int _{-i\infty}^{i\infty}\frac{d\,s}{2\pi i}\,
\left(\frac{2}{\epsilon} \right)^{n+n'-2s}\zeta
(n+n'-2s)\frac{\Gamma (s)\Gamma ^2
(\frac{n+n'+1}{2}-s)\Gamma (\frac{n+n'}{2}-s)\Gamma
(\frac{n+n'}{2}-s+1)}{\Gamma (n+1-s)\, \Gamma (n'+1-s)\,\Gamma
(n+n'+1-s)}\,,
\label{Eq38}
\end{equation}
where the integration contour is to the right of the pole at $s=0$
and to the left of the pole of $\zeta $-function at
$s=(n+n'-1)/2$. The integral is rapidly convergent as $\sim \int
ds/{\sin \pi s}$ at $s \rightarrow \pm i\infty$. For small positive $\epsilon$
the contour of integration should be enclosed around singularities
situated to the left of it. Note, that the use of the Mellin-Barnes
representation (\ref{Eq38}) for sum (\ref{Eq37}) corresponds to
chosing a definite receipt of resummation of the divergent series. 
Another possibility is to apply
to it the KLV approach~\cite{KoLiVe}. This approach was
very successful for resummation of anomalous dimensions 
obtained in three~\cite{KLOV} and four~\cite{BCDKS} loops. One can
also solve the set of linear equations (\ref{an}) approximately reducing 
it to a finite number of them with the kernel, calculated 
for all coupling constants numerically~\cite{Klebanov}. Two above methods
give similar predictions. The shortage of these approaches is that they 
do not lead to analytic results.  

One can express the kernel in terms of the sum of generalized hypergeometric
function $_4F_3$ \cite{LipPots}
\begin{equation}
K_{n,n'}(\epsilon )=\frac{\Gamma ^2(\frac{n+n'+1}{2})\,\Gamma
(\frac{n+n'}{2}+1)\,\Gamma (\frac{n+n'}{2})}{
\pi \,\Gamma (n)\,\Gamma (n'+1)\,\Gamma (n+n'+1)}\,\sum
_{k=1}^{\infty}\left(\frac{2}{k\, \epsilon} \right)^{n+n'}\,
F_{n,n'}(\frac{-4}{k^2\epsilon ^2})\,,
\label{sumhyp}
\end{equation}
where
\[
F_{n,n'}(\frac{-4}{k^2\epsilon ^2})
\]
\begin{equation}
\equiv
_4F_3(\frac{n+n'+1}{2},\frac{n+n'}{2},\frac{n+n'}{2}+1,\frac{n+n'+1}{2};
n+1,n'+1,n+n'+1;\frac{-4}{k^2\epsilon ^2})\,.
\label{hyper}
\end{equation}

\section{Iterative solution of the SE equation}

The formal solution of the ES equation for the coefficients
$a_{n,\epsilon}$ can be written as a matrix element of the
geometrical progression constructed from the operator
$\hat{K}(\epsilon)$ corresponding to the kernel $K_{n,n'}(\epsilon
)$
\begin{equation}
a_{n,\epsilon}=\sum
_{r=1}^{\infty}(-1)^{r+1}\,\left(\hat{K}^r(\epsilon)
\right)_{n,1}\,.
\end{equation}
In particular, for the anomalous dimension we have
\begin{equation}
\gamma (\epsilon )=\frac{2}{\epsilon ^2}\,\rho (\epsilon )\,,\,\,
\rho (\epsilon )=\sum _{r=0}^{\infty}(-1)^r \,
\left(\hat{K}^r(\epsilon) \right)_{1,1}\,.
\end{equation}
In two first orders one can obtain
\begin{equation}
\rho (\epsilon )=1-K_{1,1}(\epsilon )+...\,,
\end{equation}
where
\begin{equation}
K_{1,1}(\epsilon )=
\frac{1}{\pi}\int _{-i\infty}^{i\infty}\frac{d\,s}{2\pi i}\,
\left(\frac{2}{\epsilon} \right)^{2-2s}\zeta
(2-2s)\,\frac{\Gamma (s)\,\Gamma ^2
(\frac{3}{2}-s)}{(1-s)\,\Gamma (3-s)}\,.
\end{equation}
The correction $K_{1,1}(\epsilon )$ increases with decreasing
$\epsilon $. For example,
$K_{1,1}(\epsilon )\approx 8/(3\epsilon )$ for
$\epsilon \rightarrow 0$. Therefore one can obtain the solution
by the iteration in $\hat{K} (\epsilon)$ only for sufficiently
large $\epsilon$ (small $g$).

In the usual perturbation theory $1/\epsilon \rightarrow 0$ the
iteration of the solution in $\hat{K}(\epsilon )$ leads to small
corrections to $\gamma$
\[
\gamma (\epsilon)=\frac{2}{\epsilon ^2}-\frac{1}{\epsilon ^4}\zeta (2)
+\frac{1}{2\,\epsilon ^6}\,\left(3\zeta (4)+\zeta ^2(2)\right)
\]
\begin{equation}
-\frac{1}{8\,\epsilon ^8}\left(25\,\zeta (6)+12 \,\zeta (2)\,\zeta (4)
+2\zeta ^3 (2)-2\zeta ^2(3)\right) +O(\epsilon ^{-10})\,.
\end{equation}
According to the Eden-Staudacher hypothesis~\cite{EdenSt}, in the series
\begin{equation}
\gamma (\epsilon)=-8\,\sum _{k=1}^{\infty}
\left(-\frac{1}{4\epsilon ^2}\right)^{k} \,c_k\,
\label{gammaES}
\end{equation}
the quantities $c_k$ are products of $\zeta$-functions satisfying
the transcedentality principle with integer coefficients. For
example,
\begin{equation}
c_1=1\,,\,\,c_2=2 \,\zeta (2)\,,\,\,c_3=4\,\left(3\,\zeta(4)+\zeta
^2 (2)\right)\,,
\end{equation}
\begin{equation}
c_4=4\,\left(25\,\zeta (6)+12 \,\zeta (2)\,\zeta (4) +2\zeta ^3
(2)-2\zeta ^2(3)\right)\,.
\end{equation}
One can prove easily this hypothesis if the factor entering in the
expression for $K_{n,n'}(\epsilon)$
\begin{equation}
T_{n,n';R}=2n\,\frac{(2R+n+n'-1)!\,(2R+n+n')!}{R!\, (R+n)!\,
(R+n')!\,(R+n+n')!}\,.
\end{equation}
is an integer number. 

Indeed, the coefficients $c_k$ can be
presented as follows
\begin{equation}
c_k=\sum _{r=0}^{\infty}S_k^{(r)}\,,
\label{ck}
\end{equation}
where in the expression for $S_{k}^{(r)}$
\begin{equation}
S_{k}^{(r)}=\sum _{s_1=2}^{\infty}\sum _{s_2=2}^{\infty}...\sum
_{s_r=2}^{\infty}\,(-1)^{N_{s_1,s_2,...,s_r}}\,
U{s_1,s_2,...,s_r}\,
\prod _{i=1}^r \zeta
(s_i)
\label{Skr}
\end{equation}
the sum is performed over the $\zeta$-function arguments $s_t$ satisfying 
the maximal transcedentality condition 
\begin{equation}
\sum _{t=1}^rs_t=2k-2\,.
\end{equation}
It is obviously, that the number of the $\zeta$-functions with odd arguments
$s_t=2m+1$ should be even.

The factor $U{s_1,...,s_r}$ is given below
\begin{equation}
U{s_1,...,s_r}=\sum _{n_1=1}^{\infty}...\sum
_{n_{r-1}=1}^{\infty}
T_{1,n_1,\frac{s_1-1-n_1}{2}}\,
T_{n_1,n_2,\frac{s_2-n_1-n_2}{2}}...T_{n_{r-1},1,\frac{s_r-n_{r-1}-1}{2}}\,\,.
\label{Uss}
\end{equation}
Note, that the summation is performed only over such $n_l$, for
which
\begin{equation}
R_t=\frac{s_t-n_{t-1}-n_t}{2}\,
\end{equation}
is integer.
The quantities $U{s_1,...,s_r}$ are positive integer numbers providing that 
$T_{n,n';R}$ are integer. 

As for the quantity $N_{s_1,s_2,...,s_r}$ entering in the phase factor, 
it can be written as follows (see (\ref{Eq37}))
\begin{equation}
N_{s_1,s_2,...,s_r}=\sum _{t=1}^rR_t+r-(k-1)=r-1-\sum _{l=1}^{r-1}n_l\,,
\end{equation}
where we used the relation
\begin{equation}
\sum _{t=1}^rR_t=\frac{1}{2}\sum _{t=1}^rs_t-1-\sum _{s=1}^{r-1}n_s=
k-2-\sum _{s=1}^{r-1}n_s\,.
\end{equation}
The numbers $N_{s_1,s_2,...,s_r}$ are defined modulo $2$, which means,
that one can add to them an arbitrary even number. Because 
\[
2R_t=s_t-(n_{t-1}+n_t)
\]
are even numbers, one can express $N_{s_1,s_2,...,s_r}$ only
in terms of $s_l$
\begin{equation}
N_{s_1,s_2,...,s_r}=\sum _{l=1}^r(r-l)\,s_l\,.
\label{Nss}
\end{equation}
In particular, when all arguments $s_l$ of $\zeta$-functions are 
even, the phase factor $(-1)^N$ is absent. Generally it is enough
to sum only over $l$ for which $s_l$ are odd numbers. Moreover, because
such $s_l$ equal $1$ modulo $2$ one can substitute $s_l$ by $1$
\begin{equation}
N_{s_1,s_2,...,s_r}=\sum _{l}l\,\,,\,\,\,\,\,(s_l=odd)\,,
\label{Nss1}
\end{equation}
where we took into account, that the number of odd $s_l$ is
even.
 
Returning to the Eden-Staudacher hypothesis, it is helpful to present $T_{n,n';R}$ as the product 
of three factors \cite{LipPots}
\begin{equation}
T_{n,n';R}=
C_{2R+n+n'}^{R+n}\,C_{2R+n+n'}^{R}
\,\frac{2n}{2R+n+n'}\,.
\end{equation}
Here the numbers of distributions
\begin{equation}
C_{2R+n+n'}^t=\frac{(2R+n+n')!}{t!\,(2R+n+n'-t)!}
\end{equation}
are integers. The
product of all three factors is also an integer number. To
begin with, in the case when $2R+n+n'$ is a prime number $p$ the
first two factors are proportional to $2R+n+n'$ 
and can be divided
on $2R+n+n'$. If $2R+n+n'$ is a product of a prime
number $p$ and an integer number $k$, than the product of two
first factors does not contain the multiplier $p$ only if both $n$
and $n'$ are proportional to $p$, but in the last case the
multiplier $p$ is contained in the numerator of the third factor.
It proves the Eden-Staudacher hypothesis~\cite{EdenSt}.

For finite $1/\epsilon$ the corrections will be insignificant only if
the maximal eigenvalue $\lambda (\epsilon)$ of the matrix
$\hat{K}(\epsilon )$ will be sufficiently small. In an opposite
case we should use a non-perturbative approach.
For this purpose it is helpful to know the expansion
of $K_{n,n'}(\epsilon)$ near singular points $\epsilon ^2=-4$ and
$\epsilon =0$, which can be obtained from its integral
representation.

Note, that for $\epsilon ^2\rightarrow -4$ the large values of
the summation variable $R$ are essential and therefore  $\zeta (n+n'+2R)$
can be substituted by unity.  With the use of the Stirling
formulas for the $\Gamma$-functions we obtain the singularity of
$K_{n,n'}(\epsilon)$ at $\epsilon ^2=-4$ \cite{LipPots}.
\begin{equation}
K_{n,n'}(\epsilon )_{|_{\epsilon ^2\rightarrow
-4}}=\frac{n}{\pi}\,\left(\frac{2}{\epsilon}\right)^{n+n'}\,\sum
_{R=1}^{\infty}\frac{(-1)^R}{ R^2}\left(\frac{4}{\epsilon
^{2}}\right)^R\approx
\frac{n}{\pi}\,\left(\frac{2}{\epsilon}\right)^{n+n'}\,(1+\frac{4}{\epsilon
^2})\,\ln (1+\frac{4}{\epsilon ^2})\,.
\end{equation}
Using the iterative procedure near this singularity  one can
obtain the following expression for the anomalous dimension
\begin{equation}
\gamma (\epsilon )
=\frac{2/\epsilon ^2}{1+\frac{1+4/\epsilon ^2}{(1-4/\epsilon
^2)^2}\,\frac{1}{\pi}\,\ln
\left(1+4/\epsilon ^2\right)}\,.
\end{equation}
To calculate $\gamma$ at $\epsilon ^2\rightarrow -4$ with a better
accuracy we should know the regular part of $K_{n,n'}$ at  $\epsilon
^2= -4$
\[
K_{n,n'}(\pm 2\,i )=
\]
\begin{equation}
\frac{n}{\pi}\,\left(\mp i\right)^{n+n'}\sum _{R=0}^{\infty}\,
\zeta
(n+n'+2R)\,\frac{\Gamma ^2
(\frac{n+n'+1}{2}+R)\,\Gamma (\frac{n+n'}{2}+R)\,\Gamma
(\frac{n+n'}{2}+R+1)}{R!\,(R+n)!\,(R+n')!\,
(R+n+n')!}\,.
\end{equation}
The sum in this expression is convergent $\approx \sum
_{R}R^{-2}$. One can assume, that the result does not depend
essentially on $n$ and $n'$. Then the above expression for
$\gamma$ could be approximately valid at $\epsilon ^2\rightarrow
-4$ even after taking into account the regular contribution to
$K_{n,n'}$. 

Another possibility to estimate the behavior of the
kernel  $K_{n,n'}(\epsilon)$ at $\epsilon ^2\rightarrow -4$ is to
use the known expansion for the generalized hypergeometric
function $F_{n,n'} (x)$ (\ref{hyper})  for $x=-4/(k^2\epsilon ^2$) near the 
singular point
$x=1$. The summation over $k$ leads then to the singularities of
$K_{n,n'}(\epsilon)$ (\ref{sumhyp}) in the points $\epsilon ^2=-4/k^2$ for
$k=1,2,...$. Therefore at $\epsilon =0$ the kernel has an
essential singularity \cite{LipPots}. The singularities at $\epsilon ^2=-4/k^2$
appear because in the initial expression for $K_{n,n'}(\epsilon)$ (\ref{Knn})
written as an integral over $j'$ the singularities of $\phi
_{n,\epsilon }(j)$ situated at $j'=-k\,\epsilon \pm i$ pinch the
integration contour $L$ together with the square-root
singularities at $j'=\pm i$.

On the other hand, we can attempt to find the behavior of
$K_{n,n'}(\epsilon)$ for $\epsilon \rightarrow +0$ directly from
its integral representation. In this limit the integration contour
should be shifted to the right up to the first poles of $\zeta $-
and $\Gamma$- functions situated at $s=(n+n'-1)/2$ and
$s=(n+n')/2$, respectively. The contributions to $K_{n,n'}$ from
the corresponding residues are
\begin{equation}
K_{n,n'}(\epsilon )=\frac{2\,n/\epsilon }{(n+n')^2-1}\,
\frac{1}{\Gamma \left(\frac{n-n'+3}{2}\right)\,\Gamma
\left(\frac{n'-n+3}{2}\right)}-
\frac{n}{n+n'}\,\frac{1}{\Gamma
\left(\frac{n-n'+2}{2}\right)\,\Gamma
\left(\frac{n'-n+2}{2}\right)}\,.
\label{strkern}
\end{equation}

The solution of the set of linear equations for $a_{n,\epsilon}$ can
be written in the strong coupling limit $\epsilon \rightarrow 0$ as
follows
\begin{equation}
a_{n,\epsilon}=\delta _{n,1}+\epsilon \Delta a_{n,\epsilon}\,,
\end{equation}
where $\Delta a_{n,\epsilon}$ satisfies the following inhomogeneous
equation
\[
\epsilon \Delta a_{n,\epsilon}=-\delta _{n,1}-
\]
\begin{equation}
\sum
_{n'=1}^{\infty}\left(\frac{2\,n}{(n+n')^2-1}\,
\frac{1}{\Gamma \left(\frac{n-n'+3}{2}\right)\,\Gamma
\left(\frac{n'-n+3}{2}\right)}-
\frac{\epsilon \,n}{n+n'}\,\frac{1}{\Gamma
\left(\frac{n-n'+2}{2}\right)\,\Gamma
\left(\frac{n'-n+2}{2}\right)}\right)\Delta a_{n,\epsilon}\,.
\end{equation}
This set of linear equations with the leading term for $K_{n,n'}(\epsilon )$
at $\epsilon \rightarrow 0$ corresponds to the ES
equation in the form
\begin{equation}
-\epsilon \sqrt{j^2+1)}\,
\frac{\partial }{\partial j} \phi  (j)=\frac{1}{%
(j^2+1)^{1/2}+j} -2\, \int _{-i\infty}^{i\infty}\,\frac{dj^{\prime}}{2\,\pi
\,i}\sum _{n=1}^{\infty}\,\phi (j^{\prime})\,\left(\frac{-(j^{\prime
\,2}+1)^{1/2}+j^{\prime}}{(j^2+1)^{1/2}+j}\right)^n\,.
\end{equation}

The next corrections $\sim \epsilon ^k$ to the kernel
$K_{n,n'}(\epsilon )$ can be obtained from its integral
representation by taking residues in the poles situated at
$s=(n+n'+k)/2$ for $k=1,2,...$. Another possibility is to use the
known expansion of the generalized hypergeometric function
$F_{n,n'}(x)$ (\ref{hyper}), where $x=-4/(k^2\epsilon ^2)$, for the large
argument $x$.

\section{Analytic properties of the solution of the ES equation}

 By summing the geometrical progression we can
rewrite the SE equation in the form
\begin{equation}
\frac{\phi (j-\epsilon )-\phi (j)}{(j^2+1)^{-1/2}}=\frac{1}{
(j^2+1)^{1/2}+j} -2 \int _{-i\infty}^{i\infty}\,\frac{dj^{\prime}}{2\pi
i} \,\frac{\phi (j^{\prime})\,\left(-(j^{\prime
\,2}+1)^{1/2}+j^{\prime}\right)}{(j^2+1)^{1/2}+j+(j^{\prime
\,2}+1)^{1/2}-j^{\prime}}\,.
\label{ESeqj}
\end{equation}

The integration contour can be enclosed around the cut situated to the right
of it. To calculate the discontinuity one should preliminary anti-symmetrize
the integral kernel extracting from it the square-root singularity as a
factor
\[
\frac{2\,\left(-(j^{\prime \,2}+1)^{1/2}+j^{\prime}\right)}{%
(j^2+1)^{1/2}+j+(j^{\prime \,2}+1)^{1/2}-j^{\prime}}
\]
\begin{equation}
\rightarrow \frac{-(j^{\prime \,2}+1)^{1/2}+j^{\prime}}{(j^2+1)^{1/2}+j+(j^{%
\prime \,2}+1)^{1/2}-j^{\prime}}-\frac{(j^{\prime \,2}+1)^{1/2}+j^{\prime}}{%
(j^2+1)^{1/2}+j-(j^{\prime \,2}+1)^{1/2}-j^{\prime}}\,.
\end{equation}

As a result, we can write the equation in the form
\[
\left(\phi (j-\epsilon )-\phi (j)\right)\,\sqrt{j^2+1}=\frac{1}{
\sqrt{j^2+1}+j}
\]
\begin{equation}
+2\,\int _{-i}^{i}\,\frac{dj^{\prime}}{2\,\pi \,i} \,\phi
(j^{\prime})\,\left(\frac{-\sqrt{j^{\prime \,2}+1}+j^{\prime}}{%
\sqrt{j^2+1}+j+\sqrt{j^{\prime \,2}+1}-j^{\prime}}-\frac{\sqrt{j^{\prime
\,2}+1}+j^{\prime}}{\sqrt{j^2+1}+j-\sqrt{j^{\prime \,2}+1}-j^{\prime}%
}\right).
\end{equation}

It is convenient to introduce the new variable
\begin{equation}
z=(j^2+1)^{1/2}+j\,,
\label{jtoz}
\end{equation}
with the inverse relation
\begin{equation}
j=\frac{z^2-1}{2\,z}
\end{equation}
and the new function
\begin{equation}
\phi (j)=\chi (z)\,.
\end{equation}
The transformation
$z=z(j)$ performs the conformal mapping of two sheets of the
Riemann surface for $\phi (j)$ on one sheet for the function $\chi (z)$. The
physical sheet in the $j$-plane corresponds to the region $|z|>1$. The
transition to the second sheet is realized by the transformation
$z\rightarrow -z^{-1}$.

In the new variables the SE equation takes the form

\begin{equation}
\left( \chi (z_{\epsilon })-\chi (z)\right) \,\frac{z^{2}+1}{2\,z}=
\frac{1}{z}-\int_{-i}^{i}\,\frac{dz^{\prime }}{2\,\pi \,i}\,
\frac{z^{\prime 2}+1}{z^{\prime \,2}}\,
\chi (z^{\prime })\,\left( \frac{z^{\prime }}{z-z^{\prime
}}+
\frac{1/z^{\prime }}{z+1/z^{\prime }}\right) \,,
\end{equation}
where the integration over $z^{\prime }$ is done along the unit circle in
the anti-clock direction from $-i$ to $i$. We used the notation
\begin{equation}
z_{\epsilon }=\hat{R}_{\epsilon }\,z=\left( \left( \frac{z^{2}-1}{2\,z}%
-\epsilon \right) ^{2}+1\right) ^{1/2}+\frac{z^{2}-1}{2\,z}-\epsilon \,.
\label{zepsi}
\end{equation}

On the other hand, one can write the ES equation as follows
\begin{equation}
\left(\chi (z _{\epsilon})-\chi (z)\right)\,\frac{z^2+1}{2\,z}=\frac{1}{z}
-\int _{L}\,\frac{dz^{\prime}}{2\,\pi \,i}\, \frac{z^{\prime 2}+1}{z^{\prime}%
} \,\frac{\chi (\widetilde{z}^{\prime}) }{z-z^{\prime }}\,,
\end{equation}
where the integration contour $L$ chosen in a form of the unit circle is
passed in an anti-clock direction. We used the variable $\widetilde{z}$
equal to $z$ on the right part of the circle and to $-1/z$ on its left part
\begin{equation}
\widetilde{z}_{Re \,z>0}=z\,,\,\,\widetilde{z}_{Re \,z<0}=-z^{-1}.
\label{tildaz}
\end{equation}

The substitution $z\rightarrow -1/z$ in the argument of the
function $\chi (z)$ means an analytic continuation of the function
$\phi (j)$ to the second sheet of the $j$-plane with the
substitution $\sqrt{j^2+1}\rightarrow -\sqrt{j^2+1}$. But
according to the representation (\ref{Phi}) of $\phi (j)$ as a sum of the
functions $\phi _{n,\epsilon}(j)$ it has the square-root
singularities only in the points $j=-s\epsilon \pm i$ (s=1,2,...)
being  analytic at $j=\pm i$. Therefore 
\begin{equation}
\chi (\widetilde{z}')=\chi (z')\,
\label{chiztildez}
\end{equation}
and we can write the equation in the simple form
\begin{equation}
\left(\chi (z _{\epsilon})-\chi
(z)\right)\,\frac{z^2+1}{2\,z}=\frac{1}{z} -\int
_{L}\,\frac{dz^{\prime}}{2\,\pi \,i}\, \frac{z^{\prime
2}+1}{z^{\prime} } \,\frac{\chi (z^{\prime}) }{z-z^{\prime }}\,,
\label{ESKoshi}
\end{equation}
where the contour $L$ is a unit circle passed in an anti-clock direction. 
The pole at $z^{\prime
}=z$ is situated outside it. The singularities of the integral
appear when this pole and inner singularities of $\chi
(z^{\prime})$ pinch the contour. For the singular part of $\chi
(z)$ we obtain the relation
\begin{equation}
\left( \chi (z_{\epsilon })-\chi (z)\right)_{sing} \,\frac{z^{2}+1}{%
2\,z}=\frac{1}{z}-\frac{z^{2}+1}{z}\,\chi (z)_{sing}\,.
\end{equation}

Let us anti-symmetrize the right and left hand
sides of the above ES equation to the substitution $z\rightarrow
-1/z$ with the analytic continuation of the added term from $1/|z|<1$ to
$|z|=1/|z|=1$
\begin{equation}
\left(\chi (z _{\epsilon})-\chi (z)+\chi ((-1/z) _{\epsilon})-\chi
(-1/z)\right)\,\frac{z^2+1}{2\,z}=\frac{z^2+1}{z}
-\frac{z^{2}+1}{z}\,\chi (z)\,,
\label{local}
\end{equation}
where the integral contribution in its right hand side was
simplified with the use of the relations
\begin{equation}
\int _{L}\,\frac{dz^{\prime}}{2\,\pi \,i}\, \frac{z^{\prime
2}+1}{z^{\prime} } \,\left(\frac{\chi (z^{\prime}) }{z-z^{\prime
}}-\frac{\chi (z^{\prime}) }{-1/z-z^{\prime }}\right)=\int _{L}\,
\frac{dz^{\prime}}{2\,\pi \,i}\, \frac{z^{\prime 2}+1}{z^{\prime}
} \,\chi (z^{\prime})\,\left(\frac{1}{z-z^{\prime }}-\frac{z/z'
}{z-z^{\prime }}\right)\,.
\end{equation}
It is implied here, that in the first and second terms one takes
$|z|\rightarrow 1+0$
and $|z|\rightarrow 1-0$, respectively, which gives a
possibility to calculate the integral by residues taking also into
account, that
\begin{equation}
\int _{L}\,\frac{dz^{\prime}}{2\,\pi \,i}\, \frac{z^{\prime
2}+1}{z^{\prime \,2} } \,\chi (z^{\prime})=0
\end{equation}
due to the symmetry of $\chi (z^{\prime})$ under the substitution
$z^{\prime}\rightarrow -1/z^{\prime}$.

 As a result, we obtain the equation 
 \begin{equation}
\frac{1}{2}\,\left(\chi (z _{\epsilon})-\chi (z)+\chi ((-1/z)
_{\epsilon})-\chi (-1/z)\right)=1 -\chi (z)
\end{equation}
or in a simpler form \cite{LipPots}
\begin{equation}
\frac{1}{2}\,\left(\chi (z _{\epsilon})+\chi ((-1/z)
_{\epsilon})\right)=1\,. 
\label{unirel}
\end{equation}  

This equation is important, because it relates the function 
$\phi (j)$ on two sheets of the Riemann surface
\begin{equation}
\frac{1}{2}\,\left( \phi \left(j-\epsilon +\sqrt{(j-\epsilon
)^2+1}\right)+\phi \left(j-\epsilon -\sqrt{(j-\epsilon
)^2+1}\right)\right)=1\,,
\end{equation}
where $j$ takes pure imaginary values in the interval $[-i,i]$.
 In the 
$z$-representation the above relation allows one to find the 
large-$z$ asymptotics 
of the function $\chi (z)$ providing that its behavior at $z=0$
is known.

It is convenient to introduce the function
\begin{equation}
\xi (z)= \frac{z^{2}+1}{2\,z} \,\left(\chi (z _{\epsilon})-\chi
(z)\right)
\end{equation}
with the inverse relation
\begin{equation}
\chi
(z)=2\sum _{s=1}^{\infty }z_{-s \epsilon }\,\frac{\xi 
(z_{-s\epsilon })}{z^2_{-s\epsilon}+1}
=\phi (j)=\sum 
_{s=1}^{\infty }\frac{\xi \left(j+s\,\epsilon
+\sqrt{(j+s\,\epsilon )^2+1}\right)}{\sqrt{(j+s\,\epsilon )^2+1}}\,.
\label{chixi}
\end{equation}
The function $\xi (z)$ has the following simple expansion in terms of 
coefficients $a_{n,\epsilon}$ entering in ansatz (\ref{Phi}) for
$\phi (j)$
 \begin{equation}
\xi (z)=\sum _{n=1}^{\infty}z^{-n}\,
\left(\delta _{n,1}- a_{n,\epsilon }\right)
\label{xiz}
\end{equation}  
and it is analytic at $|z|>1-\delta$, where $\delta >0$.  
Moreover, this expression satisfies
the dispersion-type relation
\begin{equation}
\xi (z)= \int _{L}\,\frac{dz^{\prime}}{2\,\pi \,i}\, \frac{\xi
(z^{\prime})- \xi (-1/z^{\prime})}{z-z^{\prime }}
\label{disper}
\end{equation}
without any subtraction terms.

One can present equation (\ref{unirel}) in the form 
\begin{equation}
\xi (z)-\xi (-1/z)=\frac{z^{2}+1}{z}\,(1-\chi (z))\,.
\label{unirel1}
\end{equation}
Really the analyticity of $\xi (z)$ corresponding to Eqs. (\ref{xiz}) 
and (\ref{disper}) 
together with expression (\ref{chixi})  
for $\chi (z)$ are equivalent to the initial 
ES equation. There is an analogy of these formulas with the 
dispersion representation for the scattering amplitude and the 
unitarity condition allowing us to express its imaginary part in terms
of probabilities of various processes. In our case the expression
for  $\xi (z)-\xi (-1/z)$ is a superposition of functions $\xi (z)$ with
the shifted arguments $z\rightarrow z_{-s\epsilon}$.

In the $j$-representation Eq. (\ref{unirel1}) has the form \cite{LipPots}
\begin{equation}
\frac{\xi \left(j+\sqrt{j^2+1}\right)-\xi \left(j-\sqrt{j^2+1}
\right)}{2\,\sqrt{j^2+1}}=1-\sum _{s=1}^{\infty }\frac{\xi 
\left(j+s\,\epsilon
+\sqrt{(j+s\,\epsilon )^2+1}\right)}{\sqrt{(j+s\,\epsilon )^2+1}}
\label{unirel2}
\end{equation}
and it is equivalent to the ES equation providing that  
the analyticity condition for $\xi (z)$ at $|z|>1$ is fulfilled.

For the singularities of
$\phi (j)$  on the second sheet of the $j$-plane we obtain
\begin{equation}
\left( \phi _{sing}(j-\epsilon )-\phi _{sing}(j)\right) \,(j^{2}+1)^{1/2}=%
\frac{1}{(j^{2}+1)^{1/2}+j}-2\,\phi _{sing}(j)
\,(j^{2}+1)^{1/2}\,.
\end{equation}
Inserting in it our ansatz (\ref{Phi}) for $\phi (j)$ as a linear combination
of $\phi _{n,\epsilon }(j)$ we obtain near the corresponding
singularities
\begin{equation}
(j^2+1)^{-1/2}\,\sum
_{n=1}^{\infty}\left(j-(j^2+1)^{1/2}\right)^{n}a_{n,\epsilon
}=2\sum _{n'=1}^{\infty} \phi _{n',\epsilon}(j)\,\left(\delta
_{n',1}-a_{n',\epsilon }\right)\,.
\end{equation}
The left hand side of this equality should be analytically
continued to the second sheet. We do not discuss its singularities
at $j=\pm i$ because they are canceled exactly (see (\ref{unirel2})).
The singularities existing in
its right hand side should appear also in the left hand side as  a
result of the divergence of the sum over $n$. Thus, the analytic
properties of the ES equation lead to additional relations among
the coefficients $a_{n,\epsilon }$.

In more details, the right hand side has the square-root singularities 
situated in the points
\begin{equation}
j=-s\,\epsilon \pm i\,,\,\,s=1,2,...\,.
\end{equation}
They appear in each term of the sum over $n'$, because approximately
we have near these singularities
\begin{equation}
\frac{\left(j+s\,\epsilon+((j+s\,\epsilon )
^2+1)^{1/2}\right)^{-n'}}{((j+s\,\epsilon ) ^2+1)^{1/2}}\approx
\frac{(\pm i)^{-n'}}{(\pm 2i)^{1/2}\,(j+s \,\epsilon \mp
i)^{1/2}}\,.
\end{equation}
As a result, the condition for the appearance of the 
singularities of $\phi (j)$ on
the second sheet of the $j$-plane  at $j\rightarrow -s\,\epsilon \pm i$
has the form
\begin{equation}
\sum _{n=1}^{\infty}\left(j+(j^2+1)^{1/2}\right)^{n}a_{n,\epsilon
}=-\frac{2(s^2\,\epsilon ^2\mp 2\,i\,s\,\epsilon )^{1/2}}{(\pm
2i)^{1/2}\,(j+s \,\epsilon \mp i)^{1/2}}\,C_{\pm}(\epsilon )\,,
\end{equation}
where
\begin{equation}
j+(j^2+1)^{1/2}=\left(-s\,\epsilon \pm i+\left(s^2\,\epsilon ^2 \mp
2\,i\,s\,\epsilon \right)^{1/2}\right)\,\exp 
\left(\frac{j+s\epsilon \mp i}{\sqrt{s^2\,\epsilon ^2\mp 2\,i\,s\,\epsilon}}
\right)\,,
\end{equation}
\begin{equation}
C_{\pm}(\epsilon )=\sum _{n'=1}^{\infty}
(\pm i)^{-n'}\,\left(\delta _{n',1}-a_{n',\epsilon }\right)\,.
\end{equation}
Thus, to reproduce the singularity at $j\rightarrow -s\,\epsilon
\pm i$ the large-$n$ asymptotics of the coefficients
$a_{n,\epsilon }$ should contain the contribution
\begin{equation}
\lim _{n\rightarrow \infty}a_{n,\epsilon }\approx
-2\Re \left(\frac{\left(-s\,\epsilon + i+(s^2\,\epsilon ^2-
2\,i\,s\,\epsilon
)^{1/2}\right)^{-n}}{\sqrt{n}}\,\frac{2(s^2\,\epsilon ^2-
2\,i\,s\,\epsilon )^{1/4}}{(- 2\,\pi
\,i)^{1/2}}\,C_+(\epsilon)\right)\,.
\end{equation}
In above transformations we used the relations
\[
\lim _{x\rightarrow 1}\sum
_{n=1}^{\infty}\frac{1}{\sqrt{n}}\,x^n=\sqrt{\pi}\,(1-x)^{-1/2}
\]
and
\[
\left(1-\frac{j+(j^2+1)^{1/2}}{-s\,\epsilon \pm i+(s^2\,\epsilon
^2\mp 2\,i\,s\,\epsilon )^{1/2}}\right)^{-1/2}\approx
(s^2\,\epsilon ^2\mp 2\,i\,s\,\epsilon)^{1/4}\,(-s\,\epsilon \pm
i-j)^{-1/2}\,.
\]

The branch of the square-root in the expression $j+(j^2+1)^{1/2}$ is
defined in such way, that it grows on the second sheet at large $j$ in
all directions of
the complex plane. Therefore in the
perturbation theory $\epsilon \rightarrow \infty$ the sum over $n$ is
convergent  inside the large circle $|j|<|\epsilon|$ because for $s=1$
we have
\begin{equation}
\lim _{n\rightarrow \infty}a_{n,\epsilon }\approx
-2\Re \left(\frac{\left(-\epsilon + i-\epsilon (1-
2\,i/\epsilon
)^{1/2}\right)^{-n}}{\sqrt{n}}\,\frac{2(-\epsilon )^{1/2}\,(1-
2\,i/\epsilon )^{1/4}}{(- 2\,\pi
\,i)^{1/2}}\,C_+(\epsilon)\right)\,.
\end{equation}
In an opposite limit of the strong coupling $\epsilon \rightarrow 0$
the nearest singularity is situated at $j=\epsilon \pm i \rightarrow \pm
i$. Therefore the radius of convergence of the sum over $n$ tends to zero
for $j\rightarrow \pm i$ and $\phi
(j)$ has a singularity at $j \rightarrow \pm i$.

Note, that if one introduce the function
\begin{equation}
\Phi (j)=\phi (j-\epsilon)-\phi (j)=(j^2+1)^{-1/2}\,\sum
_{n=1}^{\infty}\left(j-(j^2+1)^{1/2}\right)^{n}\,\left(\delta
_{n,1}-a_{n,\epsilon }\right)\,,
\end{equation}
we obtain the following expression for its singular part on the second
sheet
\begin{equation}
\Phi _{sing}(j)=\frac{(j^{2}+1)^{-1/2}}{(j^{2}+1)^{1/2}+j}-2\,\sum
_{s=1}^{\infty} \Phi (j+s\,\epsilon)\,.
\end{equation}

\section{Strong coupling limit of the ES equation}

Let us consider now the strong coupling regime
\begin{equation}
\epsilon \rightarrow 0\,.
\end{equation}
To find possible corrections at large coupling constants it is helpful to use the 
expansion of $z_{\epsilon}$ (\ref{zepsi}) at small $\epsilon$
\begin{equation}
z_{\epsilon}=z-2\,\frac{z^2}{1+z^2}\,\epsilon+4\,
\frac{z^3}{(1+z^2)^3}\,\epsilon ^2+4\,
\frac{z^2(1-z^2)(1+z^4)}{(1+z^2)^5}\,\epsilon ^3+O
\left(z^3\epsilon^3\right)\,.
\end{equation}

The SE equation for $\epsilon \rightarrow 0$ is simplified 
\begin{equation}
-\epsilon \,z\,\frac{\partial }{\partial z}\,\chi (z)=\frac{1}{z} -\int
_{L}\,\frac{dz^{\prime}}{2\,\pi \,i}\, \frac{z^{\prime 2}+1}{z^{\prime}} \,%
\frac{\chi (\widetilde{z}^{\prime}) }{z-z^{\prime }}\,,
\label{ESstrong}
\end{equation}
where the integration contour is the unit circle passing
in the anti-clock direction and $\widetilde{z}$ is defined in terms of $z$ in eq. 
(\ref{tildaz}). 

The substitution $z'\rightarrow \widetilde{z}'$ in the argument
of $\chi$ is in an agreement with the linear set of equations (\ref{an})
for the coefficients of $\chi (z)$ in its expansion at
$z\rightarrow \infty$ and small $\epsilon$ (cf. (\ref{Phi}))
\begin{equation}
\chi (z)=\sum _{n=1}^{\infty}\,\frac{z^{-n}}{\epsilon
\,n}\,\left(\delta _{n,1}- a_{n,\epsilon }\right)\,.
\label{Phi1}
\end{equation}
Indeed, $a_{n,\epsilon }$ has the following representation
\begin{equation}
a_{n,\epsilon }=\int _{L}\,\frac{dz^{\prime}}{2\,\pi \,i}\,
(z^{\prime 2}+1)\,z^{\prime \,(n-2)} \, \chi
(\widetilde{z}^{\prime})=\sum
_{n'=1}^{\infty}K_{n,n'}(\epsilon)\,\left(\delta _{n',1}-  
a_{n',\epsilon }\right) \,.
\label{Eq134}
\end{equation}  
Therefore the kernel in  the strong coupling case $\epsilon \rightarrow 0$ 
is
\[
K_{n,n'}(\epsilon)=\frac{1}{\epsilon \,n'}\,\int
_{-i}^i\,\frac{dz^{\prime}}{2\,\pi \,i}\, (z^{\prime 2}+1)\,
\left(z^{\prime
\,(n-n'-2)}-(-1)^{n'}z^{\prime \,(n+n'-2)}\right)
\]
\begin{equation}
= \frac{8\,n \,\epsilon ^{-1} \,\pi ^{-1}\sin \left(\pi
\frac{n'-n-1}{2}\right)}{
\left((n+n')^2-1\right)\,\left((n-n')^2-1\right)}=  
\frac{2\,n\,\epsilon ^{-1}}{(n+n')^2-1}\, \frac{1}{\Gamma
\left(\frac{n-n'+3}{2}\right)\,\Gamma
\left(\frac{n'-n+3}{2}\right)}\,,
\label{Eq135}
\end{equation}
which is in an agreement with (\ref{strkern}).

By an anti-symmetrization of Eq. (\ref{ESstrong}) to the substitution
$z\rightarrow -1/z$ we can obtain the relation
\begin{equation}
-\epsilon \,\frac{\partial}{\partial z}
\left(\chi (z)+\chi (-1/z)\right)=\frac{1+z^2}{z} -\int
_{L}\,\frac{dz^{\prime}}{2\,\pi \,i}\, \frac{z^{\prime 2}+1}{z^{\prime}} \,%
\left(\frac{\chi (\widetilde{z}^{\prime}) }{z-z^{\prime }}-\frac{\chi (\widetilde{(-1/z)}^{\prime}) 
}{z-z^{\prime }}\right)\,,
\end{equation}
where at the integrand it is assumed, that $|z|\rightarrow 1$ in the 
first term from above 
and in the second term from below.

From relation (\ref{unirel}) valid in a general case one could obtain 
for $\epsilon \rightarrow 0$ 
\begin{equation}
-\epsilon \,\frac{z^2}{z^2+1}\,\frac{\partial}{\partial z} 
\left(\chi (z)+\chi (-1/z)\right)=1-\chi (z)\,,
\end{equation}
but this relation is valid for  $\chi$ satisfying to a different equation in comparison with 
eq. (\ref{ESstrong}). Namely, in it $\widetilde{z}'$ should be substituted by $z'$.

The singular part of the homogeneous equation for $\chi$ inside
the circle $|z|<1$ satisfies the equation
\begin{equation}
-\epsilon \,z\,\frac{\partial }{\partial z}\,\chi _{sing}^{hom}(z)= -Sing
\,\left(\frac{z^{2}+1}{z} \, \chi _{sing}^{hom}(z) \right)\,,
\end{equation}
where the sign $Sing$ means, that in the right hand side only singular terms
$\sim z^{-r}$ ($r=1,2,...$) are left.

Its solution is
\begin{equation}
\chi _{sing}^{hom}(z)=\int _{L}\,\frac{dz^{\prime}}{2\,\pi \,i}\, \frac{%
z\,\exp \frac{z^{\prime \,2}-1}{\epsilon \,z^{\prime}}}{z^{\prime
}\,(z-z^{\prime})}\,.
\end{equation}
The additional factor $z/z^{\prime}$ in the integrand leads to the constant
term $d_0$ in the large-$z$ expansion of $\chi _{sing}^{hom}$
\begin{equation}
\chi _{sing}^{hom}(z)=\sum _{k=0}^{\infty} \frac{d_k^{hom}}{z^k}\,,
\end{equation}
appearing in an agreement with singularities of the homogeneous equation.

Using
the following relation
\begin{equation}
\exp \left(\frac{z-z^{-1}}{\epsilon}\right)=\sum
_{n=-\infty}^{\infty}z^{n}\,J_n(2\,\epsilon ^{-1} )\,\,,
\,\,\,J_{-n}(x)=(-1)^n\,J_{n}(x) \,,
\end{equation}
where $J_n(x)$ are the Bessel function, we can present $\chi _{sing}^{hom}(z)$
as follows
\begin{equation}
\chi _{sing}^{hom}(z)=\sum
_{n=-\infty}^{\infty}(-z)^{-n}\,J_n(2\,\epsilon ^{-1} )\,.
\end{equation}

Let us attempt to find the singular part of the solution $\chi
(z)$ of the inhomogeneous ES equation in the form of the expansion
\begin{equation}
\chi _{sing}(z)=\sum _{k=1}^{\infty} \frac{d _k}{z^k}\,,
\label{singsolES}
\end{equation}
where the regular term with $k=0$ is absent. One can expect just such behavior 
of $\chi (z)$ at large $z$ and large coupling constants if
the string theory prediction for the anomalous dimension is right. We obtain for 
the coefficients 
$d _k$ the relations
\begin{equation}
\epsilon \,d _1=1-d _2\,
\label{DDT1}
\end{equation}
and
\begin{equation}
n\epsilon \, d _n=-d _{n-1}-d _{n+1}\,.
\label{DDT2}
\end{equation}
for $n=2,3,...$.

They are recurrent relations for the Bessel functions $J_n(2\epsilon^{-1})$
and $Y_n(2\epsilon ^{-1})$, where
\begin{equation}
Y_n(x)=\lim _{\nu \rightarrow n} \pi ^{-1} \,\left(\frac{\partial J_{\nu }(x)%
}{\partial \nu}-(-1)^n \,\frac{J_{-\nu}(x)}{\partial \nu} \right)\,.
\end{equation}
The last function has the logarithmic singularities in $x=2\epsilon ^{-1}$
due to the equality
\[
\pi \,Y_n(x)=
\]
\begin{equation}
2J_n(x)\ln (\frac{x}{2})-\sum_{m=0}^{n-1} \left(\frac{x}{2}%
\right)^{2m-n}\frac{(n-m-1)!}{m!} -\sum _{l=0}^{\infty}\left(\frac{x}{2}%
\right)^{n+2l} \frac{\psi (n+l+1)+\psi (l+1)}{l!\,(n+l)!}
\label{Ynx}
\end{equation}
incompatible with the expansion over $\epsilon ^{-1}$ in the perturbation
theory. Therefore we neglect this function and write the solution of recurrent relations 
(\ref{DDT1}), (\ref{DDT2}) only in terms of the simple Bessel function $%
J_n(2\epsilon^{-1})$ as follows
\begin{equation}
d _k=(-1)^{k+1}\,\frac{J_k(2\epsilon^{-1})}{J_0(2\epsilon^{-1})}\,.
\label{dksingES}
\end{equation}
Thus, to remove the zero mode one should impose on the solution the
constraint corresponding to the possibility of its perturbative expansion.

For the singular part of the solution at small $z$ we obtain the explicit
expression
\begin{equation}
\chi_{sing} (z)=-\frac{1}{J_{0}(2\,\epsilon ^{-1})}\,\int _L\frac{%
d\,z^{\prime}}{2\pi i}\,\frac{\exp \frac{z^{\prime 2}-1}{\epsilon
\,z^{\prime}}}{z-z^{\prime}}\,.
\label{solsingES}
\end{equation}

Note, that  $\chi _{sing}(z)$ is the solution of the equation
\begin{equation}
-\epsilon \,z\,\frac{\partial }{\partial z}\,\chi _{sing}(z)=\frac{1}{z}%
-\int_{L}\,\frac{dz^{\prime }}{2\,\pi \,i}\,\frac{z^{\prime 2}+1}{z^{\prime }%
}\,\frac{\chi _{sing}(z^{\prime })}{z-z^{\prime }}\,.
\label{ESstrong1}
\end{equation}
In comparison with eq. (\ref{ESstrong}) here $\widetilde{z}'$ is substituted by
$z'$. In principle we can obtain the equation with such substitution starting from the
ES equation (\ref{ESKoshi}) in the strong coupling limit $\epsilon \rightarrow 0$.
Thus, there is an uncertainty in the limiting form of the ES equation. Presumably it
is related to the fact, that the function $\phi (j)$ (\ref{Phi}) in the
limit $\epsilon \rightarrow 0$ develops a singularity in the point $j=\pm i$ and
therefore relation (\ref{chiztildez}) is violated. We have shown above, that from
eq. (\ref{ESstrong}) one can obtain the linear system of equations with the kernel
(\ref{Eq135}) coinciding with expression (\ref{strkern}), derived from the Mellin-Barnes
representation (\ref{Eq38}). But this representation is only one of possibilities to
sum the divergent series (\ref{Eq38}). Another way of its resummation could lead to
the kernel corresponding to eq. (\ref{ESstrong1}). At least the results obtained from
equations (\ref{ESstrong}) and (\ref{ESstrong1}) are presumably close each to another.

Having the above arguments in mind, with the use of eqs (\ref{jlim}), (\ref{anomjlim}) we obtain  
for the anomalous dimension the following result \cite{LipPots}
\begin{equation}
\gamma _{sing} =\frac{2}{\epsilon} \,\frac{J_1(2\epsilon^{-1})}{%
J_0(2\epsilon^{-1})}\,.
\label{anomsingES}
\end{equation}
In particularly in the weak coupling regime we have in an agreement with the perturbation
theory
\begin{equation}
\lim _{g\rightarrow 0}\widetilde{\gamma} \approx \frac{2}{\epsilon ^2}%
=4\,g^2\,.
\end{equation}
In the strong coupling regime this expression is simplified as follows \cite{LipPots}
\begin{equation}
\lim _{g \rightarrow \infty}\widetilde{\gamma} \approx 2\,\sqrt{2}\,g
\, \frac{\cos (\frac{2}{\epsilon} -\frac{3\pi}{4})}{\cos (\frac{2}{\epsilon}
-\frac{\pi}{4})}=2\,\sqrt{2}\,g\,
\tan \left(\frac{2}{\epsilon} -\frac{\pi}{4}\right)\,.
\label{assingES}
\end{equation}
The last result should be compared with the prediction of Polyakov and 
collaborators~\cite{GKP}
\begin{equation}
\gamma _{Pol}=2\,\sqrt{2} \, g\,.
\end{equation}
Therefore from the ES equation we obtain the anomalous dimension which in the
strong coupling limit rapidly oscillates around the string prediction.

Using the perturbative expansion of the Bessel functions we obtain in the weak
coupling limit
\begin{equation}
\gamma _{sing}=\frac{2}{\epsilon}\frac{\frac{1}{\epsilon}-\frac12 \frac{1}{%
\epsilon ^3}+\frac{1}{2!3!}\frac{1}{\epsilon ^5}}{1-\frac{1}{\epsilon ^2}+%
\frac{1}{2!2!}\frac{1}{\epsilon ^3}}=\frac{2}{\epsilon ^2}+\frac{1}{%
\epsilon^4}-\frac43 \,\frac{1}{\epsilon^6}\,.
\end{equation}
It agrees in the Born approximation $\sim 1/\epsilon ^2$ with the solution
of the SE equation, but in upper orders for the simplified version of the
equation, where the factor $t/(e^t-1)$ is substituted by unity, there are
corrections of the odd order in $1/\epsilon$. Indeed, in the leading order
\[
\chi ^{(1)}(z)=\frac{1}{\epsilon \,z}\,.
\]
The next-to-leading correction satisfies the equation
\[
-\epsilon \,z\frac{\partial }{\partial z}\,\chi ^{(2)}(z)=- \frac{1}{\epsilon%
}\int _{-i}^{i}\,\frac{dz^{\prime}}{2\,\pi \,i}\, \frac{z^{\prime 2}+1}{%
z^{\prime \,3}}\, \left(\frac{z^{\prime }}{z-z^{\prime }}+\frac{1/z^{\prime }%
}{z+1/z^{\prime }}\right)
\]
\begin{equation}
=-\frac{1}{2\pi i\,\epsilon}\left(\frac{(1+z^2)^2}{z^2}\,\ln \frac{z+i}{z-i}+%
\frac{\pi i}{z^2}+\frac{2i}{z}-2iz\right)\,.
\end{equation}
Therefore
\[
\chi ^{(2)}(z)=\frac{1}{2\pi i \,\epsilon ^2}\, \int _{\infty}^{z}
dz^{\prime}\left(\frac{(1+z^{\prime \,2})^2}{z^{\prime \,3}}\,\ln \frac{%
z^{\prime}+i}{z^{\prime}-i}+\frac{\pi i}{z^{\prime \,3}}+\frac{2i}{z^{\prime
\,2}}-2i\right)=
\]
\begin{equation}
\frac{1}{2\pi i \,\epsilon ^2}\,\left(2\int _{\infty}^z\frac{d\,z^{\prime}}{%
z^{\prime}}\,\ln \frac{z^{\prime}+i}{z^{\prime}-i}+ \frac{z^4-1}{2z^2}\,\ln
\frac{z+i}{z-i}- \frac{\pi i}{2\,z^2}-\frac{i}{z}-iz\right)\,.
\end{equation}

In particular, one can obtain the following asymptotic behavior of
$\chi ^{(2)}(z)$
\begin{equation}
\lim _{z \rightarrow \infty}
\chi ^{(2)}(z)=-\frac{16}{3\pi \,\epsilon ^{2}\,z}=-\frac{8}{3\pi
\,\epsilon ^{2}\,j}\,,
\end{equation}
which leads to the correction to the anomalous dimension
\begin{equation}
\gamma ^{(2)}=-\frac{32}{3\pi \,\epsilon ^{3}}\,.
\end{equation}
In the real case, when we do not neglect the factor $t/(e^{t}-1)$
such corrections $\sim 1/\epsilon ^{3}$ are absent.

Let us investigate now the singular behavior of $\chi (z)$ at $z
\rightarrow \pm i$. For this purpose we calculate the discontinuity
of the SE equation on the cut $-i<z<i$
\begin{equation}
\epsilon \,z\,\frac{\partial }{\partial z}\,\left(\chi (z+0)-\chi (z-0)\right)=
\frac{z^{2}+1}{z} \,\left(\chi (z+0)-\chi (-1/z)\right)\,.
\end{equation}
On the other hand, the functions $\chi (z)$ and $\chi (-1/z)$ can be
expanded near the points $z=\pm i$ in
the Taylor series. In two first orders we obtain
\begin{equation}
\chi (z+0)-\chi (-1/z)=2\chi '(\pm i) \left(z\mp i \pm \frac{i}{2}(z\mp
i)^2)\right)\,.
\end{equation}

Simplifying the right hand side of the equation near the points
$z=\pm i$ we obtain
\begin{equation}
\epsilon \,\frac{\partial }{\partial z}\,\left(\chi (z+0)-\chi (z-0)\right)
\approx \mp 4i\,\chi '(\pm i) \,(z \mp i)^2\left(1\pm
\frac{i}{2}\,(z\mp i)\right)\,.
\end{equation}
Therefore the singularities at the point $z=\pm i$ are very soft
\begin{equation}
\,\chi (z)_{z\rightarrow \pm i}=
\frac{4}{3\,\epsilon}\,\chi '(\pm i) \,
(z \mp i)^3\,\frac{\ln (z\mp i)}{2\pi }\,\left(1\pm \frac{3\,i}{4}\,(z\mp i)\right)\,.
\end{equation}
This formula is valid also in the perturbation theory $\epsilon \rightarrow \infty$,
where
\begin{equation}
\chi '(\pm i)\approx \frac{1}{\epsilon}\,.
\end{equation}

The knowledge of the singularity of the function $\chi (z)$ for $z
\rightarrow \pm i$ gives a possibility to find the asymptotic
behavior of the coefficients $a_{n,\epsilon }$ for $n \rightarrow
\infty$ in ansatz (\ref{Phi1}). Indeed, it is in an agreement with the
expansion of this function at large $z$ providing that the
coefficients $a_{n,\epsilon}$ behave at large $n$ as follows
\begin{equation}
\lim _{n \rightarrow \infty}\,a_{n,\epsilon}= \frac{8}{\pi }\,\Re
\,\chi '(i)\,\frac{i^{n+3}}{n^3} \,,
\end{equation}
where we used the relation
\begin{equation}
\lim _{x\rightarrow 1} \,\sum _{n=1}^{\infty} \frac{x^{-n}}{n^4}=
\frac{(x-1)^{3}}{3!}\,\ln (x-1)\,\,,\,\,\,x=\frac{z}{\pm i}
\end{equation}
for the singularity of the sum.

The value $\chi '(\pm i)$ can be expressed in terms of coefficients
$a_{n,\epsilon}$ as follows
\begin{equation}
\chi '(\pm i)=-\sum _{n=1}^{\infty}\,\frac{(\pm i)^{-n-1}}{\epsilon
}\,\left(\delta _{n,1}- a_{n,\epsilon }\right)\,.
\end{equation}

One can consider the ES equation on the $z$-plane, which includes
the second sheet of the $j$-plane with the cuts going from the
points $z=i$ and $z=-i$ to $z=i\infty $ and $z=-i\infty $,
respectively. We can present the function on this plane through
the dispersion integrals over these cuts and through the above
singular contribution $\chi _{sing}(z)$
\begin{equation}
\chi (z)=\int_{i}^{i\infty }\frac{dz^{\prime }}{2\pi i}\frac{\chi (z^{\prime
})-\chi (-1/z^{\prime })}{z^{\prime }-z}+\int_{-i\infty }^{-i}\frac{%
dz^{\prime }}{2\pi i}\frac{\chi (z^{\prime })-\chi (-1/z^{\prime })}{%
z^{\prime }-z}+\chi _{sing}(z)\,,
\end{equation}
where $\chi (z^{\prime })$ is defined as an analytic continuation from the
left part of the cuts. The functions $\chi (-1/z^{\prime })$ have their
argument on the line between $-i$ and $i$ where the cut is absent.

\section{Beisert-Eden-Staudacher equation}

Recently Beisert, Eden and Staudacher (BES) \cite{BES} derived
a new equation for the anomalous dimension at large $j$. In this equation 
the authors took into account the phase
factor resulting from the necessity to have an agreement with calculations
on the superstring side in the framework of the AdS/CFT correspondence \cite{BHL}. 
In Appendix A we analytically continue this phase factor to the weak coupling regime
and obtain for it the expression coinciding with that of Ref. \cite{BES}. 
Let us redefine the function $f(x)$ satisfying the ES equation (\ref{Eq1})
\begin{equation}
f(x)=\frac{t}{e^t-1}\,F(x) \,,\,\,t=\epsilon \,x
\end{equation}
and introduce the operators $K_0$ and $K_1$ acting on the new function
\begin{equation}
K_r\,F(x)=\int _0^{\infty}dx'\,\frac{t'}{e^{t'}-1}\,K_r(x,x')\,F(x')\,,
\end{equation}
where the integral kernels are (cf. (\ref{Eq5})) 
\[
K_0(x,y)= \frac{2}{x\,y}\,\sum
_{r=1}^{\infty}\,(2r-1)\,J_{2r-1}(x)\,J_{2r-1}(y)\,,\,\,
K_1(x,y)= \frac{2}{x\,y}\,\sum _{r=1}^{\infty}\,2r\,J_{2r}(x)\,J_{2r}(x)\,,
\]
\begin{equation}
K(x,y)=K_0(x,y)+K_1(x,y)\,.
\end{equation}
In particular the operators $K_0$ and $K_1$ act on the Dirac $\delta$-function
as follows
\begin{equation}
K_0\,\delta (x)=\frac{J_1(x)}{x}\,,\,\,K_1\,\delta (x)=0\,.
\end{equation}
The BES equation \cite{BES} in the operator form can be written as follows
\begin{equation}
\epsilon F(x)=\left(1+\frac{2}{\epsilon}\,K_1\right)K_0\,\delta (x)-
\left(K_0+K_1+\frac{2}{\epsilon}\,K_1\,K_0\right)F(x)\,.
\end{equation}
Note, that we leave the same definitions for $g$ and $\epsilon$ 
as in the case of the ES equation (see (\ref{def}),(\ref{def1}))
contrary to the definitions in Ref. \cite{BES}.

It is convenient  to separate the solution in its
symmetric and antisymmetric parts under the substitution 
$x\rightarrow -x$ 
\begin{equation}
F(x)=F_+(x)+F_-(x)\,,\,\,F_{\pm}(-x)=\pm F_{\pm}(x)\,.
\end{equation}
Then the BES equation is equivalent to 
the set of two equations
\[
\epsilon \,F_+(x)=K_0\,\delta (x)-
K_0\,\left(F_+(x)+F_-(x)\right)\,,
\]
\begin{equation}
\epsilon \,F_-(x)=
\frac{2}{\epsilon}\,K_1\,K_0\,\delta (x)-
K_1\left(1+\frac{2}{\epsilon}\,K_0)\right)\,\left(F_+(x)+F_-(x)\right)\,.
\label{BESpm}
\end{equation}
By finding the formal solution of the second equation for $F_-(x)$
and inserting it in the first equation for $F_+(x)$ we get
\begin{equation}
\left(1 +K_0\,\frac{1}{\epsilon +
K_1\left(1+\frac{2}{\epsilon}\,K_0\right)}\right)F_+(x)=
K_0\,\frac{1}{\epsilon +
K_1\left(1+\frac{2}{\epsilon}\,K_0\right)}\,\delta (x)\,.
\label{eqF+}
\end{equation}
Because $F_+(x)$ is even, we can present it
as follows
\begin{equation}
F_+(x)=K_+\phi (x)\,,
\end{equation}
where $\phi (x)$ is the function which does not have any symmetry under
the substitution $x\rightarrow -x$. 
From (\ref{eqF+}) we obtain the following equation for $\phi (x)$
\begin{equation}
\left(\epsilon +K_0+K_1+\frac{2}{\epsilon}\,K_1\,K_0\right)\phi (x)=
\delta (x)\,,
\end{equation}
where it was assumed, that the operator
\[
O=K_0\,\frac{1}{\epsilon +K_1+\frac{2}{\epsilon}\,K_1\,K_0}
\]
does not have zero modes.

Let us show, that one can derive the same equation for $\phi (x)$ starting from 
a simpler set of equations
\[
\epsilon \,F_+(x)=K_0\,\delta (x)-
K_0\,(F_+(x)+iF_-(x))\,,
\]
\begin{equation}
\epsilon \,F_-(x)=-K_1\,(iF_+(x)+F_-(x))\,.
\label{modBES}
\end{equation}
Indeed, by finding $F_-(x)$ from the second equation and inserting it in
the first equation we have
\begin{equation}  
\epsilon \left(1 +K_0\,\frac{1}{\epsilon 
+K_1}\,(1+\frac{2}{\epsilon}\,K_1)\right)\,F_+(x)=K_0\,\delta (x)\,.
\end{equation}
Therefore by introducing the function $\phi (x)$ as above, we obtain for it 
the same equation. Of course, the function $F_-(x)$ will be different in 
two cases, but we need only the function $F_+(x)$, because the anomalous
dimension is expressed only in terms of it
\begin{equation}
\gamma =\frac{4}{\epsilon}\,F_+(0)\,.
\end{equation}

Note, that the BES equation for $F(x)$ has the form similar to the equation 
for $\phi (x)$, but with another term in the right hand side. 
It is fulfilled, which can be verified  by adding the equations for 
$F_{\pm}(x)$.

Similar to the case of the ES equation we can construct the set of linear 
algebraic
equations equivalent to the system (\ref{modBES}) if we write the solution as 
superpositions of the Bessel functions (cf. (\ref{Phi}))
\begin{equation}
F_+(x)=\sum _{r=1}^\infty (2r-1)\frac{J_{2r-1}(x)}{x}\,
\left(\delta _{r,1}-\tilde{a}_{2r-1}\right)
\,,\,\,F_-(x)=-\sum _{r=1}^\infty (2r)\frac{J_{2r}(x)}{x}\,\tilde{a}_{2r}\,.
\end{equation}
Then the anomalous dimension is given by the same expression as earlier 
(cf. (\ref{gamma}))
\begin{equation}
\gamma =\frac{2}{\epsilon ^2}\,\left(1-\tilde{a}_{1}\right)\,,
\label{gammaBES}
\end{equation}
but for the coefficients $\tilde{a}_k$ now we have another system of  algebraic
equations (cf. (\ref{an}))
\[
\tilde{a}_{2r-1}=\sum _{r'=1}^{\infty} \,K_{2r-1,2r'-1}(\epsilon )\,
\left(\delta _{r',1}- \tilde{a}_{2r'-1}\right)-i\,\sum _{r'=1}^{\infty} 
\,K_{2r-1,2r'}(\epsilon )\,\tilde{a}_{2r'}\,,
\]
\begin{equation}
\tilde{a}_{2r}=-\sum _{r'=1}^{\infty} \,K_{2r,2r'}(\epsilon )\,
\tilde{a}_{2r'}+i\,\sum _{r'=1}^{\infty} 
\,K_{2r,2r'-1}(\epsilon )\,\left(\delta _{r',1}- \tilde{a}_{2r'-1}\right)\,,
\label{sysBES} 
\end{equation}
where the kernel $K_{n,n'}(\epsilon )$ is given by expression (\ref{Eq37}).
Therefore the coefficients of the perturbative expansion of $\gamma (\epsilon)$ 
\begin{equation}
\gamma (\epsilon)=-8\,\sum _{k=1}^{\infty}
\left(-\frac{1}{4\epsilon ^2}\right)^{k} \,\tilde{c}_k\,
\label{gammaBES}
\end{equation}
are calculated from the expression similar to (\ref{ck})
\begin{equation}
\tilde{c_k}=\sum _{r=0}^{\infty} \tilde{S}_k^{r}\,.
\end{equation}

The quantity $\tilde{S}_k^{r}$ is presented below (cf. (\ref{Skr}))
\begin{equation}
\tilde{S}_{k}^{(r)}=\sum _{s_1=2}^{\infty}\sum _{s_2=2}^{\infty}...\sum
_{s_r=2}^{\infty}\,(-1)^{\tilde{N}_{s_1,s_2,...,s_r}}\,
U{s_1,s_2,...,s_r}\,
\prod _{i=1}^r \zeta
(s_i)\,,
\label{SkrBES}
\end{equation}
where $U{s_1,s_2,...,s_r}$ are defined by expression (\ref{Uss}) and due to
the factors $i$ in coefficients of Eqs. (\ref{sysBES})
we have an additional contribution to the integer number $N_{s_1,s_2,...,s_r}$ 
presented in expression (\ref{Nss1})
\begin{equation}
\tilde{N}_{s_1,s_2,...,s_r}=\sum _{l}l+\frac{1}{2}\,\sum _{l}1\,\,\,,\,\,
\,\,s_l=2k+1\,.
\end{equation}
The additional contribution is a half of the number of $s_l$ having odd values among $r$
of them. We remind, that the number of such $s_l$ is even.

We can generalize the ES equation (\ref{ESeqj}) written for the function obtained 
by the Laplace transformation (\ref{Mellin}) to the BES case. 
We consider only the strong coupling regime, where the ES equation in the $z$-representation  
(\ref{jtoz}) is simplified (see (\ref{ESstrong})).  For the case of the BES case one should
introduce two functions satisfying the set of equations
\[
-\epsilon \,z\,\frac{\partial }{\partial z}\,\chi _+(z)=\frac{1}{z} -\frac{1}{2}\int
_{L}\,\frac{dz^{\prime}}{2\,\pi \,i}\, \frac{z^{\prime 2}+1}{z^{\prime}} \,
\left(\frac{1}{z-z^{\prime }}+\frac{1}{z+z^{\prime }}\right)\,
\left(\chi _+(\widetilde{z}^{\prime})+i\chi _-(\widetilde{z}^{\prime})\right) \,,
\]
\begin{equation}
-\epsilon \,z\,\frac{\partial }{\partial z}\,\chi _-(z)=-\frac{1}{2}\int
_{L}\,\frac{dz^{\prime}}{2\,\pi \,i}\, \frac{z^{\prime 2}+1}{z^{\prime}} \, 
\left(\frac{1}{z-z^{\prime }}-\frac{1}{z+z^{\prime }}\right)\,
\left(\chi _-(\widetilde{z}^{\prime})+i\chi _+(\widetilde{z}^{\prime})\right) \,.
\label{BESstr} 
\end{equation} 
Here the integral is performed over the unit circle in the anti-clock direction.
The variable $\widetilde{z}'$ coincides with $z'$ on the right part of the contour
$L$ and $\widetilde{z}'=-z^{\prime \,-1}$ on its left part. The variable $z$ initially
is outside the unit circle, but we analytically continue the integrals to the region
$|z|<1$. In the next section we discuss the solution of eq. (\ref{BESstr}).

\section{Strong coupling limit of the BES equation}

By simplifying  Eqs. (\ref{sysBES}) at $\epsilon \rightarrow 0$, 
we can substitute the set (\ref{DDT1}), (\ref{DDT2}) of
equations for coefficients of the singular solution (\ref{singsolES}) of the
ES equation by the corresponding equations for the BES case
\begin{equation}
\chi _{sing}^{BES}(z)=\sum _{k=1}^{\infty} \frac{\hat{d} _k}{z^k}\,,
\end{equation}
\begin{equation}
\epsilon \,\hat d _1=1-i \hat d _2\,
\label{SBES1}
\end{equation}
and
\begin{equation}
n\epsilon \, \hat d _n=-i \hat d _{n-1}- i \hat d _{n+1}\,.
\label{SBES2}
\end{equation}
for $n=2,3,...$.

They are recurrent relations for the Bessel functions $J_n(-i2\epsilon^{-1})$
and $Y_n(-i2\epsilon ^{-1})$.
But similar to the section 6, one can argue that the solution of the
above recurrent relations should contain only the Bessel functions 
$J_n(i2\epsilon^{-1})=
i^nI_n(2\epsilon^{-1})$, because the function $Y_n(i2\epsilon^{-1})$
has the singularities incompatible with the perturbation theory expansion 
(see (\ref{Ynx})). So, we obtain (cf. (\ref{dksingES}))
\begin{equation}
\hat d _k=(-i)^{k-1}\,\frac{I_k(2\epsilon^{-1})}{I_0(2\epsilon^{-1})}\,.
\end{equation}
It corresponds to the singular solution of the BES equation in the form 
(cf. (\ref{solsingES})) 
\begin{equation}
\chi_{sing}^{BES} (z)=\frac{i}{I_{0}(2\,\epsilon ^{-1})}\,\int _L\frac{%
d\,z^{\prime}}{2\pi i}\,\frac{\exp \frac{1-z^{\prime 2}}{i\,\epsilon
\,z^{\prime}}}{z-z^{\prime}}\,.
\label{solsingBES}
\end{equation}

Again, as in the case of the ES equation we stress, that $\chi_{sing}^{BES} (z)$
can be considered as a solution of the equations obtained from the strong coupling
BES equations (\ref{BESstr}) by the substitution $\widetilde{z}'\rightarrow z'$ 
(cf. (\ref{ESKoshi})). As earlier there is an uncertainty in the limiting procedure 
also for the BES equation. However we believe, that the results obtained from
the equations derived in the strong coupling limit from the different forms of the
BES equation should be close each to others.
Therefore assuming, that one can calculate the anomalous dimension from the
large-$z$ asymptotics of the
singular solution we obtain for
the anomalous dimension of the BES equation the result (cf. (\ref{anomsingES}))
\begin{equation}
\gamma _{sing}^{BES} = \frac{2}{\epsilon} \hat d _1 =
\frac{2}{\epsilon} \,\frac{I_1(2\epsilon^{-1})}{I_0(2\epsilon^{-1})}\,.
\label{anomsingBES}
\end{equation}
In the strong coupling regime $\epsilon \rightarrow 0$ the expression for 
$\gamma _{sing}^{BES}$
is significantly simplified 
\begin{equation}
\lim _{g \rightarrow \infty}\gamma _{sing}^{BES}
\approx \frac{2}{\epsilon} -\frac{1}{2} - \frac{1}{16}\,\epsilon ~=~
2\,\sqrt{2}\,g - \frac{1}{2} - \frac{\sqrt{2}}{32\,g}\,,
\label{SBES3}
\end{equation}
because the large-$t$ asymptotic of $I_{n}(t)$ is following 
\begin{equation}
I_{n}(t) = \frac{e^t}{\sqrt{2\pi t}} \left[1-\frac{4n^2 -1}{8t} 
+\frac{(4n^2 -1)(4n^2 -9)}{2!(8t)^2}
+ O\left(
\frac{1}{t^3}\right) \right]\,.
\end{equation}

The result (\ref{SBES3}) should be compared with the string prediction
\cite{GKP}
\begin{equation}
\gamma _{Pol}=2\,\sqrt{2} \, g - \frac{3}{\pi}\ln 2 
\approx 2\,\sqrt{2} \, g - 0.661907
\,.
\label{SBES4}
\end{equation}

So, we 
reproduced exactly the leading behavior for $\gamma _{Pol}$ but the NLO 
Frolov-Zeitlin coefficient is obtained only with 30\% accuracy.
Note, however, that there are additional
corrections to Eqs. (\ref{SBES1}) and (\ref{SBES2}) which 
should be responsible for the difference (see Appendix D and discussions 
therein).


\section{Conclusion}
%
In this paper we investigated the solutions of the ES and BES equations for the anomalous
dimension at the large spin $s$.
Each of these integral equations was presented as  a set of linear algebraic equations 
(\ref{an}), (\ref{sysBES}) 
with the kernel expressed in an explicit form (\ref{Eq38}), which gave us a possibility
to find the convergence radius of the small coupling expansion of this kernel and to 
calculate its strong coupling asymptotics. In particular we proved the maximal transcedentality property of
the perturbative expansion for the anomalous dimension combined with the
the ES hypothesis about the integer coefficients in front of the product of the 
$\zeta$-functions. Performing the inverse Laplace transformation of the solution together with its
subsequent conformal mapping to the $z$-plane we reduced the ES equation to the simple functional
relations (\ref{chixi}), (\ref{unirel1}) valid for all coupling constants. The similar relations are 
valid also for the solution of the BES equation. In the strong coupling limit the
singular behavior of solutions of these two equations at $z\rightarrow 0$ was constructed 
(see (\ref{solsingES}) and 
(\ref{solsingBES})). Assuming, that this singular behavior is valid also for large $z$, we calculated
the corresponding anomalous dimensions at large coupling constants (\ref{anomsingES}), (\ref{anomsingBES}).   
The asymptotic behavior of the anomalous dimension obtained from the BES equation at $g\rightarrow \infty$
(\ref{SBES3}) is in the agreement with the string predictions (see \cite{GKP}). The difference in 
corrections to these asymptotic expressions is presumably related to simplifications maid by us to get 
the exact singular solution (\ref{solsingBES})). 
Thus, we demonstrated above that the AdS/CFT correspondence together with the integrability and
transcedentality incorporated in the Beisert-Eden-Staudacher equation allow one to construct the asymptotic
behavior of the anomalous dimensions of twist-2 operators at $s\rightarrow \infty$ for $N=4$ SUSY for all 
coupling constants. However, because we do not know how to find corrections to the singular solution 
(\ref{solsingBES})) of this equation in a regular way, the problem of finding next-to-leading terms
$\epsilon \sim 1/g$ for expression (\ref{anomsingBES}) remains to be solved.

\vspace{1cm} \hspace{1cm} {\Large {} {\bf Acknowledgments} 
}

The authors are supported in part by the INTAS grant. 
L. Lipatov is thankful to the Hamburg and Paris-6 Universities for
their hospitality during the period of time when this work was completed. He
was supported also by the Maria Curie and RFBR grants.  

We are indebted to 
B. Eden, V. Kazakov, R. Janik, A. Onishchenko, A. Rej, D. Serban, M. Staudacher and
V. Velizhanin 
for helpful discussions.

\newpage
\section*{Appendix A. Dressing phase}
 \label{App:A}
\def\theequation{A\arabic{equation}}
\setcounter{equation}0

The dressing phase factor according to Refs. \cite{BES,GoHe}
should have the following form
\begin{equation}
\exp(i\theta_{12})=
\exp\left[i\sum_{r=2}^{\infty} \sum_{s=r+1}^{\infty} c_{r,s}
\left( q_r\left(x_1^{\pm}\right) q_s\left(x_2^{\pm}\right)-
q_s\left(x_1^{\pm}\right) q_r\left(x_2^{\pm}\right)\right)\right].
\label{AA1}
\end{equation}
Here
$q_r(x)$ are the conserved magnon charges
\begin{equation}
 q_r\left(x^{\pm}\right)=\frac{i}{r-1}\left(\frac{1}{(x^{+})^{r-1}}
- \frac{1}{(x^{+})^{r-1}} \right)
\label{AA2}
\end{equation}
and $c_{r,s}$ are some coefficients depending on the coupling constant 
$g=\sqrt{\lambda}/4\pi$, where $\lambda$ is the 't Hooft coupling. 

The strong coupling expansion of $c_{r,s}$ was obtained 
on the string theory side \cite{ArFS,BeTs,BHL}
\begin{equation}
c_{r,s}=\frac{\left(1-(-1)^{r+s}\right)}{2} \, (r-1)(s-1) \, 
\hat c_{r,s},~~~~
\hat c_{r,s} =  \sum_{n=0}^{\infty} \hat c^{(n)}_{r,s} g^{1-n}.
\label{AA3}
\end{equation}
Here the coefficients
\begin{eqnarray}
\hat c^{(n)}_{r,s}= \frac{1}{(-2\pi)^n} \, \frac{\zeta(n)}{\Gamma(n-1)} \,
\frac{\Gamma\left[\frac{1}{2}(s+r+n-3)\right]
\Gamma\left[\frac{1}{2}(s-r+n-1)\right]}{
\Gamma\left[\frac{1}{2}(s+r-n+1)\right]
\Gamma\left[\frac{1}{2}(s-r-n+3)\right]}
\label{AA4}
\end{eqnarray}
were derived in \cite{BHL} with the use of the crossing symmetry \cite{Janik}.

Recently the week coupling expression for $c_{r,s}$ was suggested in \cite{BES}
in the form of the expansion
\begin{equation}
\hat c_{r,s} =  -\sum_{n=0}^{\infty} \hat c^{(-n)}_{r,s} g^{1+n},
\label{AA5}
\end{equation}

The purpose of this appendix is to prove expression (\ref{AA5}).

\subsection*{A.1 Basic formulas}

Eq.(\ref{AA3}) means that only the odd integer values of the sum $r+s$ 
are important, i.e.
\begin{equation}
r+s=2m+1, 
\label{AA6}
\end{equation}
where $m\geq 2$, because $r\geq 2$ and $s\geq 3$.

Further, following to Ref. \cite{BES} it is useful to extract from the sum in
(\ref{AA2}) the first two terms:
\begin{equation}
\hat c_{r,s} =  \hat c^{(1)}_{r,s} \, g + \hat c^{(0)}_{r,s} + B_{r,m},
\label{AA6}
\end{equation}
where
\begin{eqnarray}
 c^{(0)}_{r,s} &=& \frac{\delta_{r+1,s}}{(r-1)r} = \frac{\delta_{r,m}}{(r-1)r},
\label{AA7}\\
 c^{(1)}_{r,s} &=& -\frac{2}{\pi} \, \frac{1}{(s+r-2)(s-r)} = -\frac{2}{\pi} \,
\frac{1}{(2m-1)(2(m-r)+1)},
\label{AA8}\\
B_{r,m} &=& \sum_{l=0}^{\infty} \hat B^{(l)}_{r,m} g^{-(l+1)}
\label{AA8.1}
\end{eqnarray}
and
\begin{eqnarray}
B^{(l)}_{r,m} = 
\frac{1}{(-2\pi)^{l+2}} \, \frac{\zeta(l+2)}{\Gamma(l+1)} \,
\frac{\Gamma(m+l/2) \Gamma(m+1-r+l/2)}{\Gamma(m-l/2) \Gamma(m+1-r-l/2)}
\label{AA9}
\end{eqnarray}


Let us consider now eq. (\ref{AA5}). According to \cite{BES} 
$\hat c^{(n)}_{r,s}$ 
can be presented  using the relation
\begin{eqnarray}
\zeta(1-z)=2(2\pi)^{-z} \cos\left(\frac{\pi}{2}z\right) \Gamma(z) \zeta(z),~~
\Gamma(1-z)= \frac{\pi}{\sin(\pi z)  \Gamma(z)}
\label{AA10}
\end{eqnarray}
in the form (for odd values of $r+s$)
\begin{eqnarray}
\hat c^{(n)}_{r,s}=
\frac{2 (-1)^{s-1-n}\cos\left(\frac{\pi}{2}n\right)\zeta(1-n)\Gamma(2-n)
\Gamma(1-n)}{
\Gamma\left[\frac{1}{2}(5-n-r-s)\right]
\Gamma\left[\frac{1}{2}(3-n+r-s)\right]
\Gamma\left[\frac{1}{2}(3-n-r+s)\right]
\Gamma\left[\frac{1}{2}(1-n+r+s)\right]}
\nonumber
\end{eqnarray}

For $r+s=2m+1$ and $n=2k$ (
$\cos[\pi(k+1/2)]=0$ and, thus,
$\hat c^{(-2k-1)}_{r,s}=0$)
the eq. (\ref{AA5}) 
can be rewritten as follows
\begin{equation}
\hat c_{r,s} =  -2 \sum_{k=0}^{\infty} (-1)^{k+r}
\frac{\zeta(1+2k)\Gamma(2+2k)
\Gamma(1+2k)}{
\Gamma(2-m+k)\Gamma(r+1-m+k)\Gamma(2+m-r+k)\Gamma(m+1+k)}\, g^{1+2k}
\label{AA12}
\end{equation}
For $k \leq m-2$ the coefficients are zero and, thus, one can substitute
$\sum_{k=0}^{\infty} \to \sum_{k=m-1}^{\infty}$ in the r.h.s. .
Then, eq. (\ref{AA12}) can be rewritten in the new variable $p=k-m+1$
\begin{eqnarray}
\hat c_{r,s} &=&  2 \sum_{p=0}^{\infty} (-1)^{m+r+p}
\frac{\zeta(2p+2m-1)\Gamma(2p+2m-1)\Gamma(2p+2m)}{
\Gamma(p+1)\Gamma(p+r)\Gamma(p+2m+1-r)\Gamma(p+2m)}\, g^{2p+2m-1}
\nonumber \\
&=& 2 (-1)^{m+r} \, g^{2m-1} \, \hat K_{r-1,2m-r},
\label{AA13}
\end{eqnarray}
where
\begin{equation}
 \hat K_{r,s}= \sum_{p=0}^{\infty} (-g^2)^{p}
\frac{\zeta(2p+r+s)(2p+r+s)!(2p+r+s-1)!}{
p!(p+r)!(p+s)!(p+r+s)!}.
\label{AA14}
\end{equation}

Note that $\hat K_{r,s}$ contributes to the perturbative results for
the ES and BES equations (see \cite{LipPots,BES}), because
\begin{equation}
 \hat K_{r,s}= \int^{\infty}_{0} dz \, \frac{J_r(2gz)J_s(2gz)}{z(e^z-1)} 
\label{AA15}
\end{equation}
and $J_r(2gz)$ is the Bessel function.

\subsection*{A.2 Results}

Let us write the Mellin-Barnes representation for the coefficients $B_{r,m}$
\begin{eqnarray}
B_{r,m} = \int_{-i\infty}^{i\infty} \frac{dt}{2\pi i}
\frac{1}{(2\pi)^{2-t}} \, \Gamma(t) \zeta(2-t)\,
\frac{\Gamma(m-t/2) \Gamma(m+1-r-t/2)}{\Gamma(m+t/2) \Gamma(m+1-r+t/2)}
\, g^{t-1},
\label{AB1}
\end{eqnarray}
where the integration contour is to the right of the pole at $t=0$ and
to the left of the pole at $t=1$.
Closing the  contour 
to the left,
we can reconstruct the
Eqs. (\ref{AA8.1}) and (\ref{AA9}) from the poles of $ \Gamma(t)$ at 
$t\to -l$.

Using relation (\ref{AA10}) for $\zeta(1-z)$ (\ref{AA10}) one can rewrite
(\ref{AB1}) as follows
\begin{eqnarray}
B_{r,m} &=& \frac{1}{\pi} \int_{-i\infty}^{i\infty} \frac{dt}{2\pi i}
 \, \Gamma(t) \cos\left(\frac{\pi}{2}(t-1)\right) \Gamma(t-1)
\zeta(t-1) \nonumber \\
&&
\, 
\frac{\Gamma(m-t/2+\Delta) \Gamma(m+1-r-t/2+\delta)}{\Gamma(m+t/2) 
\Gamma(m+1-r+t/2)}
\, g^{t-1},
\label{AB2}
\end{eqnarray}
where small quantities $\Delta$ and $\delta$ were added to the argument of 
the last above 
$\Gamma$-function to prevent contributions from coinciding poles.

Closing now the  integration contour of eq.(\ref{AB2})
to the right,
we have four terms 
coming from the poles of $\Gamma(t-1)$, $\zeta(t-1)$, $\Gamma(m-t/2+\Delta)$ and
$\Gamma(m+1-r-t/2+\delta)$, respectively.\\

{\bf 1.}~~ The pole of $\Gamma(t-1)$ at $t \to 1$ gives at $\Delta=\delta=0$
(with $\zeta(0)=-1/2$)
\begin{eqnarray}
 {}_1B_{r,m} ~=~  \frac{1}{2\pi} \,
\frac{1}{(m-1/2)(m-r+1/2)},
\label{AB3}
\end{eqnarray}
and cancels the term $c^{(1)}_{r,s}$ in eq. (\ref{AA6}).\\

{\bf 2.}~~ The pole of $\zeta(t-1)$ at $t =2+2\varepsilon$ produces
at $\Delta=\delta=0$
\begin{eqnarray}
 {}_2B_{r,m} ~=~-
\frac{1}{\pi} 
 \,  \cos\left(\frac{\pi}{2}(1+2\varepsilon)\right) 
\Gamma(2+2\varepsilon) \Gamma(1+2\varepsilon)
\,
\frac{\Gamma(m-1-\varepsilon) \Gamma(m-r-\varepsilon)}{\Gamma(m+1+\varepsilon) 
\Gamma(m+2-r+\varepsilon)}
\, g^{1+2\varepsilon}\,.
\nonumber
\end{eqnarray}

The result is zero if $r\neq m$, because
\begin{eqnarray}
 \cos\left(\frac{\pi}{2}(1+2\varepsilon)\right) ~=~ - \pi \varepsilon \,.
\label{AB5}
\end{eqnarray}

For $r=m$ there is an additional pole at $\varepsilon \to 0$ in
$\Gamma(-\varepsilon)$. So, finally we have
\begin{eqnarray}
 {}_2B_{r,m} ~=~ - \frac{\delta_{r,m}}{m(m-1)} \, g \,,
\label{AB6}
\end{eqnarray}
which cancels the term $c^{(0)}_{r,s} \, g$ in eq. (\ref{AA6}).\\

{\bf 3.}~~ $\Gamma(m-t/2+\Delta)$ produces poles
at $t \to 2(m+k+\Delta)$. One can obtain at $\delta=0$
\begin{eqnarray}
 &&{}_3B_{r,m} ~=~
\frac{1}{\pi} \sum_{k=0}^{\infty} \, \frac{(-1)^k}{k!} 
 \,  \cos\left[\pi\left(m+k+\Delta -\frac{1}{2}\right)\right] 
\Gamma(2(m+k+\Delta)-1) 
\nonumber \\
&& \zeta(2(m+k+\Delta)-1) 
\,
\frac{\Gamma(2(m+k+\Delta)) \Gamma(1-k-r-\Delta)}{\Gamma(2m+k+\Delta) 
\Gamma(2m+k+1-r+\Delta)}
\, g^{2(m+k+\Delta)-1}.
\label{AB7}
\end{eqnarray}

Because
\begin{eqnarray}
&&\cos\left[\pi\left(m+k+\Delta -\frac{1}{2}\right)\right]~=~
(-1)^{m+k} \pi \Delta + O\left( \Delta^2 \right), \nonumber \\
&&\Gamma(1-k-r-\Delta) ~=~ (-1)^{r+k}  \frac{\Gamma(1-\Delta)\Gamma(\Delta)}{
\Gamma(k+r+\Delta)}, 
\label{AB8}
\end{eqnarray}
eq. (\ref{AB7}) can be rewritten as follows
\begin{eqnarray}
 {}_3B_{r,m} &=&
\sum_{k=0}^{\infty} \, (-1)^{m+r+k} 
 \, 
\frac{\Gamma(2(m+k)) \Gamma(2(m+k)-1) \zeta(2(m+k)-1)}{k! \Gamma(k+r)
\Gamma(k+2m) 
\Gamma(k+2m+1-r)}
\, g^{2(m+k)-1}
\nonumber \\ 
&=& (-1)^{m+r} \, g^{2m-1} \, 
\hat K_{r-1,2m-r},
\label{AB9}
\end{eqnarray}
i.e.
${}_3B_{r,m}$ is two times less then $\hat c_{r,s}$ (see eq. (\ref{AA13})).\\

{\bf 4.}~~ $\Gamma(m+1-r-t/2+\delta)$ produces poles
at $t \to 2(m+1-r+k+\delta)$. 

Similar to the previous subsection after simple algebraic transformations
one can obtain the contribution 
${}_4B_{r,m}$, which coincides with 
${}_3B_{r,m}$, i.e.
\begin{eqnarray}
 {}_4B_{r,m} ~=~ {}_3B_{r,m} ~=~ (-1)^{m+r} \, g^{2m-1} \, 
\hat K_{r-1,2m-r}\,.
\label{AB10}
\end{eqnarray}

So, the sum ${}_3B_{r,m} + {}_4B_{r,m}$ reproduces the result (\ref{A13})
\begin{eqnarray}
 {}_3B_{r,m}+ {}_4B_{r,m} ~=~ 
2 (-1)^{m+r} \, g^{2m-1} \, 
\hat K_{r-1,2m-r} ~=~ \hat c_{r,s}\,.
\label{AB11}
\end{eqnarray}

Thus, we proved the duality between large and week expansions of
of $c_{r,s}$ at odd values of $s+r$
\begin{equation}
\hat c_{r,s} =  \sum_{n=0}^{\infty} \hat c^{(n)}_{r,s} g^{1-n} =
- \sum_{n=0}^{\infty} \hat c^{(-n)}_{r,s} g^{1+n},~~~(s+r=2m+1)
\label{AC1}
\end{equation}
with $c^{(n)}_{r,s}$ given by Eq. (\ref{AA4}).

\newpage
\section*{Appendix B. Independent derivation of linear algebraic equations}
 \label{App:B}
\def\theequation{B\arabic{equation}}
\setcounter{equation}0

Here we present the derivation of the set of linear equations (\ref{an})
without using the Laplace transformation (\ref{Mellin}).

{\bf 1.}~~ We start with the initial version  (\ref{Eq1}) of the ES equation
with the kernel  (\ref{Eq5}) and write it as follows
\begin{equation}
\Psi(x)= \frac{J_1(x)}{x}- 2 \sum_{k=1}^{\infty} \, k \frac{J_k(x)}{x} \,
\int _0^{\infty}
\frac {dz}{e^{\epsilon z}-1}  J_k(z) \Psi(z),
\label{A1}
\end{equation}
where the new function
\begin{equation}
\Psi(x)=\frac{e^{\epsilon x}-1}{x} \, f(x)\,.
\label{A1.5}
\end{equation}
is introduced.

One can search its solution  in the form
\begin{equation}
\Psi(x)= 
\sum_{n=1}^{\infty} \hat a_{n} \frac{J_{n}(x)}{x}
\label{A2}
\end{equation}

By inserting the ansatz (\ref{A2}) in Eq. (\ref{A1}) and comparing the 
coefficients  in the front of $J_k(x)$ in the l.h.s. and r.h.s. , we obtain
\begin{equation}
\hat a_{k}=\delta_{k,1} - 2k b_{k},
\label{A3}
\end{equation}
where
\begin{eqnarray}
b_{k}&=& \int _0^{\infty}
\frac {dz}{e^{\epsilon z}-1}  J_k(z) \Psi(z) \nonumber \\
&=& \frac{1}{\epsilon} \, \int _0^{\infty}
\frac {dt}{e^{t}-1}  J_k(t/\epsilon) \Psi(t/\epsilon) 
= \sum_{n=1}^{\infty} \, \frac{1}{\epsilon} \, \int _0^{\infty}
\frac {dt}{e^{t}-1}  J_k(t/\epsilon) 
 d_n \frac{J_{n}(t/\epsilon)}{(t/\epsilon)}\,.
\label{A4}
\end{eqnarray}
Eqs. (\ref{A3}) and (\ref{A4}) are similar to  (\ref{an}) 
and (\ref{Knn}) in the main text.

Expanding both Bessel functions in the r.h.s. of Eq. (\ref{A4}) in series 
according to  (\ref{Eq3}) and applying the integral representation for the
Euler $\zeta$-function
\begin{equation}
p! \zeta(p+1)=  \, \int _0^{\infty}
\frac {dt}{e^{t}-1} \, t^p,
\label{A5}
\end{equation}
we obtain the following expression for $b_{k}$
\begin{equation}
b_{k}= 
\sum_{n=1}^{\infty} \hat a_n \sum_{s=0}^{\infty} \frac{1}{s!(s+k)!}
 \sum_{l=0}^{\infty} \frac{(-1)^{s+l}}{l!(l+n)!}
\, \Bigl(2(s+l)+k+n-1\Bigr)! \zeta\Bigl(2(s+l)+k+n\Bigr) \, 
\overline{g}^{2(s+l)+k+n},
\label{A6}
\end{equation}
where $\overline{g}=g/\sqrt{2}=1/(2\epsilon)$.

The interchange of the summation order  gives
\begin{equation}
b_{k}= 
\sum_{n=1}^{\infty} \hat a_n \overline{g}^{k+n}
 \sum_{p=0}^{\infty} \, (-\overline{g}^2)^{p}
\, \Bigl(2p+k+n-1\Bigr)! \zeta\Bigl(2p+k+n\Bigr) \,
\sum_{s=0}^{\infty} \frac{1}{s!(s+k)!(p-s)!(p-s+n)!}.
\label{A6.1}
\end{equation}

The last term in the r.h.s. can be calculated exactly
\begin{eqnarray}
\sum_{s=0}^{\infty} \frac{1}{s!(s+k)!(p-s)!(p-s+n)!}
&=& \frac{{}_2F_1(-p,-(p+n),k+1,1)}{s!(s+k)!(p-s)!(p-s+n)!} 
\nonumber \\
&=&
\frac{\Bigl(2p+k+n\Bigr)!}{p!(p+k)!(p+n)!(p+k+n)!},
\label{A6.2}
\end{eqnarray}
where ${}_2F_1$ is the hypergeometric function.

Thus, we obtain
\begin{eqnarray}
b_{k}&=& 
\sum_{n=1}^{\infty} \hat a_n \, \overline{g}^{k+n} \,
\sum_{p=0}^{\infty} 
\frac{(2p+n+k)!(2p+n+k-1)!\zeta(2p+k+n)}{p!(p+n)!(p+k)!(p+n+k)!}
 \, (-\overline{g}^2)^p \nonumber \\
&\equiv & \frac{1}{2k} \,  
\sum_{n=1}^{\infty} \hat a_n \, \overline{g}^{k+n} \, K_{k,n}(\overline{g}),
\label{A7}
\end{eqnarray}
where $ K_{k,n}$ is given by Eq. (\ref{Eq37}).

So, the obtained set of equations is
\begin{equation}
\hat a_{k}=\delta_{1,k} - 
\sum_{n=1}^{\infty} \hat a_n \, \overline{g}^{k+n} \, K_{k,n}(\overline{g}) ,
\label{A8}
\end{equation}
which coincides with Eqs (\ref{Phi}) 
and (\ref{an}), if $\hat a_n \to a_{n,\epsilon}$. The coincidence is
not trivial, because  $\hat a_n$ and  $a_{n,\epsilon}$ are
coefficients of the expansions of the functions $\Psi(x)$ and $f(x)$, respectively.


\vspace{0.5cm}

{\bf 2.}~~ Let us consider now the BES equation (see Eq. (B3) in [BES]) in the form
similar to (\ref{A1})
\begin{eqnarray}
\Psi(x)&=& \frac{J_1(x)}{x} + 2 \sum_{k=1}^{\infty} (-1)^k 
c_{2k+1,2}(\overline{g}) \, \frac{J_{2k}(x)}{x} \,
- 2 \sum_{k=1}^{\infty} \, k \frac{J_k(x)}{x} \,
\int _0^{\infty}
\frac {dz}{e^{\epsilon z}-1}  J_k(z) \Psi(z)
\nonumber \\
&-& 4 \sum_{k=1}^{\infty} \, \frac{J_{2k}(x)}{x} \,
\sum_{s=1}^{\infty} (-1)^{k+s+1} 
c_{2k+1,2s}(\overline{g}) \,
\int _0^{\infty}
\frac {dz}{e^{\epsilon z}-1}  J_{2s-1}(z) \Psi(z),
\label{A10}
\end{eqnarray}
where
\begin{eqnarray}
c_{r,s}(\overline{g}) &=& 2 \cos\left[\frac{\pi}{2} (s-r-1)\right]
(r-1)(s-1)\, 
\int _0^{\infty}
\frac {dt}{t(e^{t}-1)}  J_{r-1}(\overline{g}t)  J_{s-1}(\overline{g}t)
\nonumber \\
&\equiv& 2 \cos\left[\frac{\pi}{2} (s-r-1)\right]
(r-1)(s-1)\, \widetilde{c}_{r,s}(\overline{g})
\label{A11}
\end{eqnarray}

One can introduce even and odd components of $\Psi(x)$, such that
 $\Psi_{\pm}(-x)=\pm \Psi_{\pm}(x)$:
\begin{equation}
\Psi(x)= \Psi_{+}(x)+\Psi_{-}(x),~~ 
\Psi_{\pm}(x)=\frac{1}{2} \Bigl(\Psi(x)\pm \Psi(-x)\Bigr)
\label{A12}
\end{equation}

Adding and subtracting
Eq. (\ref{A10}) and that  with the replacement
$x \to -x$, we have
\begin{eqnarray}
\Psi_+(x)&=& \frac{J_1(x)}{x} 
- 2 \sum_{k=1}^{\infty} \, (2k-1) \frac{J_{2k-1}(x)}{x} \,
\int _0^{\infty}
\frac {dz}{e^{\epsilon z}-1}  J_{2k-1}(z) \Psi(z)
\label{A13} \\
\Psi_-(x)&=& 2 \sum_{k=1}^{\infty} (-1)^k 
c_{2k+1,2}(\overline{g}) \, \frac{J_{2k}(x)}{x} \,
- 2 \sum_{k=1}^{\infty} \, (2k) \frac{J_{2k}(x)}{x} \,
\int _0^{\infty}
\frac {dz}{e^{\epsilon z}-1}  J_{2k}(z) \Psi(z)
\nonumber \\
&-& 4 \sum_{k=1}^{\infty} \, \frac{J_{2k}(x)}{x} \,
\sum_{s=1}^{\infty} (-1)^{k+s+1} 
c_{2k+1,2s}(\overline{g}) \,
\int _0^{\infty}
\frac {dz}{e^{\epsilon z}-1}  J_{2s-1}(z) \Psi(z).
\label{A14}
\end{eqnarray}

The solutions for $\Psi_{\pm}(x)$ can be found by analogy with the 
previous subsection. They have the form
\begin{eqnarray}
\hat a_{2k-1}&=&\delta_{k,1} - 2(2k-1) b_{2k-1}, \label{A15}\\
\hat a_{2k}&=& 2(-1)^k c_{2k+1,2}(\overline{g})
\delta_{k,1} - 4k b_{2k} - 4 \sum_{s=1}^{\infty} (-1)^{k+s+1} 
c_{2k+1,2s}(\overline{g}) \, b_{2s-1}\,.
\label{A16}
\end{eqnarray}

Thus, for $\hat a_{2k-1}$ we have
\begin{equation}
\hat a_{2k-1}=\delta_{1,k} - 
\sum_{n=1}^{\infty} \hat a_n \, \overline{g}^{2k+n-1} \, K_{2k-1,n}(\overline{g})\, .
\label{A17}
\end{equation}

To obtain the expression for $\hat a_{2k}$ we should calculate 
$c_{2k+1,2s}(\overline{g})$.
Expanding both Bessel functions in the r.h.s. of Eq. (\ref{A11}) in series 
according to  (\ref{Eq3}) and applying the integral representation  (\ref{A5})
for the Euler $\zeta$-functions,
we obtain 
\begin{eqnarray}
\widetilde{c}_{2k+1,2s}(\overline{g})&=& 
\sum_{m=0}^{\infty}  \frac{1}{m!(m+2k)!}  
 \sum_{l=0}^{\infty} \frac{(-1)^{m+l}}{l!(l+2s-1)!}
\, \Bigl(2(k+s+m+l-l)\Bigr)! \nonumber \\
&\cdot&
\zeta\Bigl(2(k+s+m+l)-1\Bigr) \, 
\overline{g}^{2(k+s+m+l)-1},
\label{A18}
\end{eqnarray}

Analogously to Eqs. (\ref{A6.1}) and (\ref{A6.2}) we interchange the 
summation order 
\begin{eqnarray}
\widetilde{c}_{2k+1,2s}(\overline{g})&=& 
\sum_{p=0}^{\infty} (-1)^p \, \overline{g}^{2(p+k+s)-1} \,
\frac{\Bigl[(2p+s+k-1)\Bigr]!\Bigl[(2p+s+k)-1\Bigr]!}{p!(p+2s-1)!
(p+2k)!(p+2s+2k-1)!}
\nonumber \\
&\cdot&
\zeta(2p+k+s)-1)
~\equiv ~  \overline{g}^{2(k+s)-1} \,
\frac{1}{4k} \,   K_{2k,2s-1}(\overline{g})\,,
\label{A19}
\end{eqnarray}
and obtain
\begin{eqnarray}
c_{2k+1,2s}(\overline{g}) = \frac{2s-1}{2} (-1)^{s+k} 
\overline{g}^{2(k+s)-1} \,  K_{2k,2s-1}(\overline{g})\,,
\label{A20}
\end{eqnarray}
which is in agreement with eq. (\ref{AA13}).

Using Eqs. (\ref{A8}) and (\ref{A20}), one obtains
\begin{eqnarray} 
\hat a_{2k}&=& 
\overline{g}^{2k+1} \,  K_{2k,1}(\overline{g}) -
\sum_{n=1}^{\infty} \hat a_n \, \overline{g}^{2k+n} \, K_{2k,n}(\overline{g})
\nonumber \\
&+& 
\sum_{n=1}^{\infty} \hat a_n \, \overline{g}^{2k+n} \,
 \sum_{s=1}^{\infty} \overline{g}^{2s-2} \,
\, K_{2k,2s-1}(\overline{g}) \, K_{2s-1,n}(\overline{g})
\label{A21}
\end{eqnarray}

Note that in agreement with (\ref{sysBES}), the eqs. (\ref{A17}) and 
(\ref{A21}) can be rewritten in more simpler form
\begin{eqnarray}
\hat a_{2k-1}&=&\delta_{1,k} - 
\sum_{m=1}^{\infty} \hat a_{2m-1} \, \overline{g}^{2k+2m-2} \, 
K_{2k-1,2m-1}(\overline{g}) - i 
\sum_{m=1}^{\infty} \hat a_{2m} \, \overline{g}^{2k+2m-1} \, 
K_{2k-1,2m}(\overline{g}),
\nonumber \\
\hat a_{2k}&=& -i 
\sum_{m=1}^{\infty} \hat a_{2m-1} \, \overline{g}^{2k+2m-1} \, 
K_{2k,2m-1}(\overline{g}) - 
\sum_{m=1}^{\infty} \hat a_{2m} \, \overline{g}^{2k+2m} \, 
K_{2k,2m}(\overline{g}),
\label{A23}
\end{eqnarray}
which coincides with the one for ES equation with replacement
$K_{r,s} \to i K_{r,s}$ for the odd values of the sum $r+s$.

\section*{Appendix C. Linear algebraic equations for $\epsilon \to 0$}
 \label{App:C}
\def\theequation{C\arabic{equation}}
\setcounter{equation}0

Here we derive the set of linear algebraic equations for solution of 
the ES equation at large coupling constant limit $g \to \infty$.

The equations in this limit can be obtained from the Mellin-Barnes 
representation (\ref{Eq38}) for the kernel $K_{n n'}$.
 Nevertheless it is useful to derive them independently.

At $\epsilon \to 0$, Eq. (\ref{Eq1}) is simplified as follows
\begin{equation}
\epsilon \,f(x)=\frac{J_1(x)}{x}-
2 \sum_{k=1}^{\infty} \, k \frac{J_k(x)}{x} \,
\int _0^{\infty}
\frac {dz}{z}  J_k(z) f(z)
\label{B1}
\end{equation}

Similar to Appendix B
one can introduce even and odd components of $f(x)$, such that
 $f_{\pm}(-x)=\pm f_{\pm}(x)$:
\begin{equation}
f(x)= f_{+}(x)+f_{-}(x),~~ f_{\pm}(x)=\frac{1}{2} \Bigl(f(x)\pm f(-x)\Bigr)
\label{B2}
\end{equation}

Adding and subtracting
Eq. (\ref{B1}) and that  with the replacement
$x \to -x$, we have
\begin{eqnarray}
\epsilon \,f_+(x)&=&\frac{J_1(x)}{x}-
2 \sum_{k=1}^{\infty} \,(2k-1) \frac{J_{2k-1}(x)}{x} \,
\int _0^{\infty}
\frac {dz}{z}  J_{2k-1}(z) f(z)
\label{B3}\\
\epsilon \,f_-(x)&=& -4 \sum_{k=1}^{\infty} \, k \frac{J_{2k}(x)}{x} \,
\int _0^{\infty}
\frac {dz}{z}  J_{2k}(z) f(z)
\label{B4}
\end{eqnarray}

\vspace{0.5cm}

{\bf 1.} We can present the function $f(x)$ in the form similar to (\ref{A2})
\begin{equation}
f(x)= \sum_{n=0}^{\infty} a_{n+1} \frac{J_{n+1}(x)}{x}, ~~
f_+(x)= \sum_{n=0}^{\infty} a_{2n+1} \frac{J_{2n+1}(x)}{x}, ~~
f_-(x)= \sum_{n=0}^{\infty} a_{2n+2} \frac{J_{2n+2}(x)}{x}.
\label{B5}
\end{equation}

The integrals appearing in the r.h.s. of (\ref{B3}) and
(\ref{B4}) can be calculated as follows
\begin{eqnarray}
\int _0^{\infty} \frac {dz}{z^2}  J_{2s}(z)  J_{2l}(z) &=&
\frac{4}{\pi} 
\frac{(-1)^{s+l+1}}{\left[4(s+l)^2-1\right]\left[4(s-l)^2-1\right]}
\label{B6} \\
\int _0^{\infty} \frac {dz}{z^2}  J_{2s+1}(z)  J_{2l+1}(z) &=&
\frac{4}{\pi} 
\frac{(-1)^{s+l+1}}{\left[4(s+l+1)^2-1\right]\left[4(s-l)^2-1\right]}
\label{B7} \\
\int_0^{\infty} \frac{dz}{z^2}  J_{2s}(z)  J_{2l+1}(z) &=&
\frac{1}{4} 
\frac{\Gamma(s+l)}{\Gamma(s+l+2)\Gamma(1+s-l)\Gamma(l-s+2)} 
\nonumber \\
&& 
= \left\{\begin{array}{rl}
  \frac{\delta_s^l}{8s(2s+1)}, & \mbox{if}~ s=l , \\
  \frac{\delta_s^{l+1}}{8s(2s-1)}, & \mbox{if}~ s=l+1 \ .
 \end{array} \right.
\label{B8} 
\end{eqnarray}

Therefore we obtain the following linear set of equations
for coefficients $a_k$  
\begin{eqnarray}
\epsilon \, a_{2s+1}&=& \delta_{0,s} - \frac{a_{2s}(1-\delta_{0,s})}{4s}
-  \frac{a_{2(s+1)}}{4(s+1)}
\nonumber \\
&& -  \frac{8(2s+1)}{\pi} \, \sum_{m=0}^{\infty} \, 
\frac{(-1)^{m+s+1} a_{2m+1}}{\left[4(m+s+1)^2-1\right]\left[4(m-s)^2-1\right]}
\label{B10} \\
\epsilon \, a_{2s+2}&=&
- \frac{a_{2s+1}}{2(2s+1)}
-  \frac{a_{2s+3}}{2(2s+3)}
\nonumber \\
&& -  \frac{16(s+1)}{\pi} \, 
\sum_{m=0}^{\infty} \, 
\frac{(-1)^{m+s+1} a_{2m+2}}{\left[4(m+s+2)^2-1\right]\left[4(m-s)^2-1\right]}
\label{B11}
\end{eqnarray}

The first and second equations came 
from Eqs. (\ref{B3}) and (\ref{B4}), respectively.
These results can be derived also directly form
Eqs. (\ref{Eq134}) and (\ref{Eq135}) with replacement 
$a_{n,\epsilon} \to a_{2s+1}$ , i.e. $n \to 2s+1$.

For the singular form of ES equation the formulas can be obtained by
the replacement $a_p \to 2 a_p$ ($p=2s,2s+1,2s+2,2s+3$) and $a_{2m+1}= a_{2m+2}=0$
in the r.h.s. of (\ref{B10}) and (\ref{B11}).
 So, the equations are similar for odd and 
even part and can be rewritten in the simpler form
\begin{eqnarray}
\epsilon \, a_{n}&=& \delta_{1,n} - \frac{a_{n-1}(1-\delta_{1,n})}{(n-1)}
-  \frac{a_{n+1}}{(n+1)},
\label{B12}
\end{eqnarray}
which coincides with the ones in (\ref{DDT1}) and (\ref{DDT2})  
after the replacement $a_n=n d_n$.\\

{\bf 2.} We can search the functions $f(x)$,  $f_+(x)$ and $f_-(x)$
in another form
\begin{equation}
f(x)= \sum_{n=0}^{\infty} \tilde d_n J_{n}(x), ~~
f_+(x)= \sum_{n=0}^{\infty} \tilde d_{2n} J_{2n}(x), ~~
f_-(x)= \sum_{n=1}^{\infty} \tilde d_{2n-1} J_{2n-1}(x).
\label{B13}
\end{equation}

By substituting this ansatz in Eqs. (\ref{B3}) and (\ref{B4}) and
using the formulas
\begin{eqnarray}
\int _0^{\infty} \frac {dz}{z}  J_{2s}(z)  J_{2l}(z) &=&
\frac{1}{4s} \, \delta_{s.l} 
\label{B14} \\
\int _0^{\infty} \frac {dz}{z}  J_{2s+1}(z)  J_{2l+1}(z) &=&
\frac{1}{2(2s-1)} \, \delta_{s,l} 
\label{B15} \\
\int_0^{\infty} \frac{dz}{z}  J_{2s}(z)  J_{2l-1}(z) &=& 
\frac{2(-1)^{s+l}}{\pi [4s^2-(2l-1)^2]}
\label{B16} 
\end{eqnarray}
we obtain the following set of equations for coefficients $d_k$ 
\begin{eqnarray}
2 (2l-1) \epsilon
\,  \sum_{m=0}^{l-1} \tilde d_{2m}
+ O\Bigl(\epsilon^2 \Bigr) &=&
\delta_{l,1} -  \tilde d_{2l-1} - 
\frac{4 (2l-1)}{\pi}
\sum_{m=0}^{\infty} \, 
\frac{(-1)^{m+l} \tilde d_{2m}}{\left[4m^2-(2l-1)^2\right]},
\label{B17} \\
4l \epsilon
\, \sum_{m=0}^{l-1} \tilde d_{2m+1}
+ O\Bigl(\epsilon^2 \Bigr) &=&
- \tilde d_{2l} -
\frac{8l}{\pi}
\sum_{m=0}^{\infty} \, 
\frac{(-1)^{m+l} \tilde d_{2m-1}}{\left[4l^2-(2m-1)^2\right]} 
~~~~(l \geq 1),
\label{B18}
\end{eqnarray}
derived from the comparison of coefficients in the front of the 
functions
$J_{2l-1}(x)/x$ and $J_{2l}(x)/x$, respectively. The equations
(\ref{B17}) and (\ref{B18}) look simpler then ones 
(\ref{B10}) and (\ref{B12}).\\

{\bf 3.} Eqs. (\ref{B17}) and (\ref{B18})  can be simplified by 
considering as independent ones the coefficients $\tilde d_{2l-1}$.
Indeed, from Eq. (\ref{B18}) one can express the coefficients $\tilde d_{2l}$
through $\tilde d_{2l-1}$ and insert  the results in the l.h.s. of Eq. (\ref{B17}).

Using some algebraic manipulations we obtain at $\epsilon =0$
\begin{eqnarray}
\tilde d_{0} &=&
 \frac{\pi}{4} \delta_{l,1} + 
\frac{(-1)^{l}(2l-1)}{2\pi} \,  \tilde d_{2l-1} \,
\Bigl(\Psi'(l+1/2)
+ \frac{2}{(2l-1)^2} \Bigr)  \nonumber \\
&-& \frac{(2l-1)^2}{2\pi} \left[
\sum_{m=1}^{l-1} + \sum_{m=l+1}^{\infty} \right]
\, 
(-1)^{m} \tilde d_{2m-1} R(l,m),
\label{B19}
\end{eqnarray}
where
\begin{eqnarray}
R(l,m) &=& \frac{1}{2(l-m)(l+m-1)} \,
\biggl[2  \Psi(m+1/2) - \pi tg(\pi m) -  \Psi(l+1/2) + \pi tg(\pi l)
\nonumber \\
&&+\frac{4(l-m)}{(1-2m)(1-2l)} \biggr] ~= ~ \frac{1}{(l-m)(l+m-1)} \,
\biggl[\Psi(m+1/2) -  \Psi(l+1/2) \biggr] 
\nonumber \\
&&- \frac{2}{(1-2m)(1-2l)(l+m-1)}
\label{B20}
\end{eqnarray}
and
$\Psi$ and $\Psi'$ are Riemann $\Psi$-function and its derivation.
The r.h.s. result is correct for integer $m$ and $l$ values.

We see that now the number of equations is two times less in comparison 
with the Section B.2, but there are more complicated coefficients in the 
front of the numbers $\tilde d_{2l-1}$.
Note that if one would express all odd coefficients $\tilde d_{2l-1}$ through even ones
$\tilde d_{2l}$, the formulas look even more complicated.

The linear equations (\ref{B10}), (\ref{B12}), (\ref{B17}),  (\ref{B18})
and (\ref{B19})  
can be solved by the matrix equation technique (see \cite{Tracy} and 
references therein).
We plan to return to the consideration of these equations in future.

\section*{Appendix D. Linear algebraic equations for BES equation at
$\epsilon \to 0$}
 \label{App:D}
\def\theequation{D\arabic{equation}}
\setcounter{equation}0

Here we derive the set of linear algebraic equations from 
the BES equation at large coupling constants $g \to \infty$.

Let us 
introduce even and odd components of $f(x)$ as in eq. (\ref{B2}).

According to above considerations, the set of equations for $f_{\pm}(x)$
in the framework of the BES equation can be obtained directly 
from (\ref{B3}) and (\ref{B4}) by the replacement $f_{-}(z) \to
if_{-}(z)$ in the r.h.s. of (\ref{B3}) and $f_{+}(z) \to
if_{+}(z)$ in the r.h.s. of (\ref{B4}):
\begin{eqnarray}
\epsilon \,f_+(x)&=&\frac{J_1(x)}{x}-
2 \sum_{k=1}^{\infty} \,(2k-1) \frac{J_{2k-1}(x)}{x} \,
\int _0^{\infty}
\frac {dz}{z}  J_{2k-1}(z) \Bigl( f_+(z)+i f_-(z)\Bigr)
\label{D1}\\
\epsilon \,f_-(x)&=& -4 \sum_{k=1}^{\infty} \, k \frac{J_{2k}(x)}{x} \,
\int _0^{\infty}
\frac {dz}{z}  J_{2k}(z) \Bigl( i f_+(z)+ f_-(z)\Bigr).
\label{D2}
\end{eqnarray}

\vspace{0.5cm}

{\bf 1.} Let start with the form (\ref{B13})
for the functions $f(x)$,  $f_+(x)$ and 
$f_-(x)$ considered in the subsection {\bf 2} of the Appendix C.

Following the above argument and in an agreement with (\ref{D1}) and
(\ref{D2}), to have the set of equations for 
coefficients $\tilde d_k$ in the case of the BES equation we should replace
$\tilde d_{2l-1} \to i \tilde d_{2l-1}$ and $\tilde d_{2l} \to i \tilde d_{2l}$ in Eqs. 
(\ref{B17}) and (\ref{B18}), respectively. So, one can obtain
the following linear algebraic equations
\begin{eqnarray}
2 (2l-1) \epsilon
\,  \sum_{m=0}^{l-1} \tilde d_{2m}
+ O\Bigl(\epsilon^2 \Bigr) &=&
\delta_{l,1} -  i \tilde d_{2l-1} - 
\frac{4 (2l-1)}{\pi}
\sum_{m=0}^{\infty} \, 
\frac{(-1)^{m+l} \tilde d_{2m}}{\left[4m^2-(2l-1)^2\right]},
\label{D3} \\
4l \epsilon
\, \sum_{m=0}^{l-1} \tilde d_{2m+1}
+ O\Bigl(\epsilon^2 \Bigr) &=&
- i \tilde d_{2l} -
\frac{8l}{\pi}
\sum_{m=0}^{\infty} \, 
\frac{(-1)^{m+l} \tilde d_{2m-1}}{\left[4l^2-(2m-1)^2\right]}\,,
\label{D4}
\end{eqnarray}
derived from the comparison of coefficients in the front of the functions
$J_{2l-1}(x)/x$ and $J_{2l}(x)/x$, respectively. \\

{\bf 2.} We can present also the function $f(x)$ in the form (\ref{B5}). 
Using integrals (\ref{B6})-(\ref{B8}), one can obtain
 the following linear set of equations
for coefficients $a_k$  
\begin{eqnarray}
\epsilon \, a_{2s+1}&=& \delta_{0,s} - i \frac{a_{2s}(1-\delta_{0,s})}{4s}
-  i \frac{a_{2(s+1)}}{4(s+1)}
\nonumber \\
&& -  \frac{8(2s+1)}{\pi} \, \sum_{m=0}^{\infty} \, \frac{16}{\pi} \,
\frac{(-1)^{m+s+1} a_{2m+1}}{\left[4(m+s+1)^2-1\right]\left[4(m-s)^2-1\right]}
\label{D5} \\
\epsilon \, a_{2s+2}&=&
- i \frac{a_{2s+1}}{2(2s+1)}
-  i \frac{a_{2s+3}}{2(2s+3)}
\nonumber \\
&& -  \frac{16(s+1)}{\pi} \, 
\sum_{m=0}^{\infty} \, \frac{16}{\pi} \,
\frac{(-1)^{m+s+1} a_{2m+2}}{\left[4(m+s+2)^2-1\right]\left[4(m-s)^2-1\right]},
\label{D6}
\end{eqnarray}
which coincides with the one in Eqs. (\ref{B10}) and (\ref{B11}) with
the replacement $a_{2s} \to i a_{2s}$ and $a_{2(s+1)} \to i a_{2(s+1)}$
in the r.h.s. of (\ref{B10}) and
$a_{2s+1} \to i a_{2s+1}$ and $a_{2s+3} \to i a_{2s+3}$
in the r.h.s. of (\ref{B11}), respectively.


By analogy with the
Appendix B we can show, that the results for the singular form of BES equation 
can be obtained by 
the replacement $a_p \to 2 a_p$ ($p=2s,2s+1,2s+2,2s+3$) and $a_{2m+1}= a_{2m+2}=0$
in the r.h.s of (\ref{D5}) and (\ref{D6}). So, 
the equations should be similar for odd and 
even part and can be rewritten in simpler form
\begin{eqnarray}
\epsilon \, a_{n}&=& \delta_{1,n} - i \frac{a_{n-1}(1-\delta_{1,n})}{(n-1)}
-  i \frac{a_{n+1}}{(n+1)},
\label{D7}
\end{eqnarray}
which coincides with the ones in (\ref{SBES1}) and (\ref{SBES2}) after the
 replacement $a_n=n \hat d_n$.
These equations have been already solved in the Section 8.

Note that the solution has corrections of two types. The first ones
came from the last terms in the r.h.s of (\ref{D5}) and (\ref{D6})
and should be responsible for the difference between $\epsilon$-corrections
in eq. (\ref{SBES3}) and the corresponding Frolov-Tseytlin correction
(see (\ref{SBES4})).

The second type of corrections to results for anomalous dimension in
(\ref{SBES3}) comes from the corrections to the l.h.s of (\ref{D5}) and (
\ref{D6}). Indeed, here we keep only the first term at $\epsilon \to 0$
of the factor $(e^{\epsilon x}-1)/x \cdot f(x)$. Expanding $ e^{\epsilon x}$,
by analogy with (\ref{B17}) and (\ref{B18})
we obtain the additional contributions in the l.h.s of (\ref{D5}) 
and (\ref{D6}):
\begin{eqnarray}
\epsilon \, a_{2s+1} &\to& \epsilon \, a_{2s+1} + \epsilon^2
\, (2s+1) \Bigl(1-\delta_{0,s}\Bigr) \sum_{m=0}^{s-1} a_{2m+2}
+ O\Bigl(\epsilon^3 \Bigr),
\label{D8} \\
\epsilon \, a_{2s+2} &\to& \epsilon \, a_{2s+2} + \epsilon^2
\, (2s+2) \sum_{m=0}^{s} a_{2m+1}
+ O\Bigl(\epsilon^3 \Bigr)
\label{D9}
\end{eqnarray}

The study of the both type of the corrections is the subject of our future
investigations.


\begin{thebibliography}{0}

\bibitem{DGLAP}
V.~N.~Gribov and L.~N.~Lipatov, Sov.\ J.\ Nucl.\ Phys.\ \textbf{15} (1972) 438;
V.~N.~Gribov and L.~N.~Lipatov, Sov.\ J.\ Nucl.\ Phys.\ \textbf{15} (1972) 675;
L.~N.~Lipatov, Sov.\ J.\ Nucl.\ Phys.\ \textbf{20} (1975) 94;
G.~Altarelli and G.~Parisi, Nucl.\ Phys.\ \textbf{B126} (1977) 298;
Yu.~L. Dokshitzer, Sov.\ Phys.\ JETP \textbf{46} (1977) 641.

\bibitem{LONLOAD}  D.~J.~Gross and F.~Wilczek, Phys.\ Rev.\ D \textbf{8}
(1973) 3633; 
H.~Georgi and H.~D.~Politzer, Phys.\ Rev.\ D \textbf{9} (1974)
416; 
E.~G.~Floratos, D.~A.~Ross and C.~T.~Sachrajda, Nucl.\ Phys.\
\textbf{B129} (1977) 66; [Erratum-ibid.\ \textbf{B139} (1978) 545]; E.
G.~Floratos, D.~A.~Ross and C.~T.~Sachrajda, Nucl.\ Phys.\ \textbf{B152}
(1979) 493; 
A.~Gonzalez-Arroyo, C.~Lopez and F.~J.~Yndurain, Nucl.\ Phys.\
\textbf{B153} (1979) 161; 
A.~Gonzalez-Arroyo and C.~Lopez, Nucl.\ Phys.\
\textbf{B166} (1980) 429;
G.~Gurci, W.~Furmanski and R.~Petronzio, Nucl.\ Phys.\
\textbf{B175} (1980) 27; 
W.~Furmanski and R.~Petronzio, Phys.\ Lett.\
\textbf{B97} (1980) 437;
E.~G.~Floratos, C.~Kounnas and R.~Lacage, Nucl.\ Phys.\
\textbf{B192} (1981) 417; 
C.~Lopes and F.~J.~Yndurain, Nucl.\ Phys.\
\textbf{B171} (1980) 231, 
\textbf{B183} (1981) 157;
R.~Hamberg and W.~L.~van Neerven, Nucl.\ Phys.\ \textbf{B379} (1992) 143;
R.~K.~Ellis and W.~Vogelsang, arXiv:hep-ph/9602356;
R.~Mertig and W.~L.~van Neerven, Z.\ Phys.\ \textbf{C70}
(1996) 637; 
W.~Vogelsang, Nucl.\ Phys.\ \textbf{B475} (1996) 47.
%
\bibitem{VMV}
S.~Moch, J.~A.~M.~Vermaseren and A.~Vogt,
Nucl.\ Phys.\ B {\bf 688} (2004) 101;
 A.~Vogt, S.~Moch and J.~A.~M.~Vermaseren,
Nucl.\ Phys.\ B {\bf 691} (2004) 129.

\bibitem{KL}
A.~V.~Kotikov and L.~N.~Lipatov, Nucl.\ Phys.\ \textbf{B661} (2003) 19;
arXiv:hep-ph/0112346.

\bibitem{KoLiVe}
A.~V.~Kotikov, L.~N.~Lipatov and V.~N.~Velizhanin,
Phys.\ Lett.\ B \textbf{557} (2003) 114.

\bibitem{BFKL}
L.~N.~Lipatov, Sov.\ J.\ Nucl.\ Phys.\ \textbf{23} (1976) 338;
V.~S.~Fadin, E.~A.~Kuraev and L.~N.~Lipatov, Phys.\ Lett.\ B \textbf{60} (1975) 50;
E.~A.~Kuraev, L.~N.~Lipatov and V.~S.~Fadin, Sov.\ Phys.\ JETP \textbf{44} (1976) 443;
E.~A.~Kuraev, L.~N.~Lipatov and V.~S.~Fadin, Sov.\ Phys.\ JETP \textbf{45} (1977) 199;
I.~I.~Balitsky and L.~N.~Lipatov, Sov.\ J.\ Nucl.\ Phys.\ \textbf{28} (1978) 822;
I.~I.~Balitsky and L.~N.~Lipatov, JETP\ Lett.\ \textbf{30} (1979) 355.

\bibitem{next}
V.~S.~Fadin and L.~N.~Lipatov, Phys.\ Lett.\ B \textbf{429} (1998) 127;
G.~Camici and M.~Ciafaloni, Phys.\ Lett.\ B \textbf{430} (1998) 349.

\bibitem{KL00}
A.~V.~Kotikov and L.~N.~Lipatov, Nucl.\ Phys.\ \textbf{B582} (2000) 19.

\bibitem{KLOV}
A.~V.~Kotikov, L.~N.~Lipatov, A.~I.~Onishchenko and V.~N.~Velizhanin,
Phys.\ Lett.\ B {\bf 595} (2004) 521; (E: B {\bf 632} (2006) 754).

\bibitem{Konish} B.~Eden, C.~Jarczak and E.~Sokatchev, 
Nucl. Phys. \ B {\bf 712}
(2005) 157.

\bibitem{Bern1} Z.~Bern, L.J.~Dixon, V.A.~Smirnov, Phys. Rev. \ D {\bf 72} 
(2005) 085001.

\bibitem{BCDKS} Z. Bern, M. Czakon, L. Dixon, D. Kosover and V. Smirnov,
arXiv: hep-th/0610248.

\bibitem{AdS-CFT}  J.~Maldacena, Adv.\ Theor.\ Math.\ Phys.\ \textbf{2}
(1998) 231;
 Int.\ J.\ Theor.\ Phys.\ \textbf{38} (1998) 1113;
S.~S.~Gubser, I.~R.~Klebanov and A.~M.~Polyakov, Phys.\
Lett.\ B \textbf{428} (1998) 105; 
E.~Witten, Adv.\ Theor.\ Math.\ Phys.\
\textbf{2} (1998) 253.

\bibitem{BKP} J. ~Bartels, Nucl.\ Phys.\ B {\bf 175} (1980) 413;
J. Kwiecinski and M. Praszalowich, Phys.\ Lett.\ B {94} (1980) 413. 

\bibitem{integrBF} L. ~N. ~Lipatov, Phys.\ Lett.\ B {\bf 309} (1993)
394; 
{\it High energy asymptotics of multi-colour QCD and 
exactly solvable lattice models}, Padova preprint DFPD/93/TH/70,
hep-th/9311037, unpublished.

\bibitem{Heis}
L.~N.~Lipatov,
JETP Lett.\ \textbf{59} (1994) 596; 
L.~D.~Faddeev and G.~P.~Korchemsky,
Phys.\ Lett.\ B \textbf{342} (1995) 311.

\bibitem{QPO}  A.\ P.\ Bukhvostov, E.\ A.\ Kuraev, L.\ N.\ Lipatov and
G.\ V.\ Frolov,
JETP Lett.\ \textbf{41} (1985) 92;
A.~P.~Bukhvostov, G.~V.~Frolov, L.~N.~Lipatov and E.~A.~Kuraev,
Nucl.\ Phys.\ \textbf{B258} (1985) 601.

\bibitem{integrDG}  L.N.~Lipatov, Perspectives in Hadronic Physics, in:
\textit{Proc. of the ICTP conf.} (World Scientific, Singapore, 1997).
Nucl. Phys. Proc. Suppl. \textbf{99A} (2001) 175.

\bibitem{BDMKB} V.M.~Braun, S.E.~Derkachev, A.N.~Manashov, Phys. Rev. Lett.
{\bf 81} (1998) 2020; 
V.M.~Braun, S.E.~Derkachev, G.P. Korchemsky, A.N.~Manashov, 
Nucl.\ Phys.\ B {\bf 553} (1999) 355; 
A.V. Belitsky, Phys.\ Lett.\ B {\bf 453}
(1999) 59.

\bibitem{BeiSta} Niklas Beisert and Matthias Staudacher, Nucl.\ Phys.\
\textbf{B727} (2005) 1.   





\bibitem{EdenSt}
B.~Eden and M.~Staudacher, {\it Integrability and Transcedentality},
arXiv:hep-th/0603157.

\bibitem{Bel}
A.V. Belitsky, {\it  Long-range SL(2) Baxter equation in N=4 
super-Yang-Mills theory}, arXiv:hep-th/0609068.

\bibitem{BES} N. Beisert, B. Eden and M. Staudacher, 
{\it Transcedentality and Crossing},
arXiv:hep-th/0610251.

\bibitem{LipPots} L.N. Lipatov, 
{\it Transcendality and the Eden-Staudacher equation}, talk at
the Workshop on Integrability in Gauge and String Theory, AEI, Potsdam,
Germany, July 24-28, 2006.


\bibitem{Klebanov} M.K. Benna, S. Benvenuti, I.R. Klebanov and A. Scardicchio,
arXiv: hep-th/0611135

\bibitem{GKP} S. Gubser, I. Klebanov and A. Polyakov, Nucl.\ Phys.\ B {\bf 638}
(2002) 99;
S. Frolov and A.A. Tseytlin, JHEP {\bf 0206} (2002) 007.




\bibitem{BHL} N. Beisert, R. Hernandez and E. Lopez, 
{\it A Crossing-Symmetric Phase
for $AdS_5 - S^5$ Strings}, hep-th/0609044.



\bibitem{GoHe}  N. Beisert and T. Klose, J. Stat. Mech. \textbf{0607} (2006) 
P006; 
C. Gomez and R. Hernandez,  arXiv:hep-th/0611014.

\bibitem{ArFS}
G.~Arutyunov, S. Frolov and M.~Staudacher, JHEP \textbf{0410} (2004) 016.

\bibitem{BeTs}
N. Beisert and A.A. Tseytlin, Phys.\ Lett.\ B {\bf 629} (2005) 102;
 R. Hernandez and E. Lopez, JHEP \textbf{0607} (2006) 004;
L. Freyhult and C. Kristjansen, Phys.\ Lett.\ B {\bf 638} (2006) 258.


\bibitem{Janik}
R.A. Janik, Phys.\ Rev.\ D {\bf 73} (2006) 086006.

\bibitem{Tracy} C.A. Tracy and H. Widom, arXiv: hep-th/9210073.

\end{thebibliography}
\end{document}